\documentclass[12pt]{article}
\usepackage{epsfig, amsmath, amssymb}
\usepackage{hyperref}
\usepackage{graphicx,psfrag}
\usepackage{xcolor}
\usepackage{slashed,verbatim,subfigure}
\usepackage{appendix}
\catcode`@11
\def\seceqaa{\@addtoreset{equation}{section}
	\def\theequation{A\arabic{equation}}}
\def\seceqbb{\@addtoreset{equation}{section}
	\def\theequation{B\arabic{equation}}}
	\def\seceqcc{\@addtoreset{equation}{section}
	\def\theequation{C\arabic{equation}}}

\newcommand{\nn}{\nonumber}
\newcommand{\be}{\begin{equation}}
\newcommand{\ee}{\end{equation}}
\newcommand{\ben}{\begin{equation}}
\newcommand{\een}{\end{equation}}
\newcommand{\bea}{\begin{eqnarray}}
\newcommand{\eea}{\end{eqnarray}}
\newcommand{\bA}{\begin{array}}
\newcommand{\eA}{\end{array}}
\newcommand{\bc}{\begin{center}}
\newcommand{\ec}{\end{center}}
\newcommand{\al}{\alpha}

\newcommand{\ra}{\rightarrow}
\newcommand{\del}{\partial}

\newcommand{\ie}{{\it i.e.}}
\newcommand{\eg}{{\it e.g.}}

\numberwithin{equation}{section}

\begin{document}


\begin{titlepage}

%

\bc

\hfill 
\\         [25mm]

{\Huge Cosmological singularities, \\ [2mm]
 holographic complexity and entanglement} 
\vspace{16mm}

{\large K.~Narayan,\ \ Hitesh K. Saini,\ \ Gopal Yadav} \\
\vspace{3mm}
{\small \it Chennai Mathematical Institute, \\}
{\small \it H1 SIPCOT IT Park, Siruseri 603103, India.\\}

\ec
\vspace{35mm}

\begin{abstract}
  We study holographic volume complexity for various families of
  holographic cosmologies with Kasner-like singularities, in
  particular with $AdS$, hyperscaling violating and Lifshitz
  asymptotics. We find through extensive numerical studies that the
  complexity surface always bends in the direction away from the
  singularity and transitions from spacelike near the boundary to
  lightlike in the interior. As the boundary anchoring time slice
  approaches the singularity, the transition to lightlike is more
  rapid, with the spacelike part shrinking. The complexity functional
  has vanishing contributions from the lightlike region so in the
  vicinity of the singularity, complexity is vanishingly small,
  indicating a dual Kasner state of vanishingly low complexity,
  suggesting an extreme thinning of the effective degrees of freedom
  dual to the near singularity region. We also develop further
  previous studies on extremal surfaces for holographic entanglement
  entropy, and find that in the IR limit they reveal similar behaviour
  as complexity.
\end{abstract}


\end{titlepage}

{\tiny 
\begin{tableofcontents}    
\end{tableofcontents}
}


\vspace{-2mm}

\section{Introduction}

Among the various quantum information ideas and tools that have become
ubiquitous in holography over the last several years, a fascinating
class of questions involves computational complexity. Complexity
measures the difficulty in preparation of the final state from some
initial reference state. Discussions of eternal black holes dual to
thermofield double states and ER=EPR \cite{Maldacena:2013xja}
suggested that the linear growth in time of the spatial volume of the
bulk Einstein-Rosen bridge is dual to the linear time growth of
complexity in the dual field theory
\cite{Susskind:2014rva,Stanford:2014jda,Susskind:2014jwa,Roberts:2014isa,Susskind:2014moa}. This is encapsulated in the expression
\be \label{CV}
C(t) \sim {{\rm Vol}(\Sigma_t)\over G_N R}\,,
\ee
for complexity $C(t)$ in terms of an extremal codim-1 spacelike
slice $\Sigma_t$ at anchoring time $t$, with $G_N$ the Newton constant
in $AdS$ with scale $R$. The precise proportionality factors are not
canonical to pin down and are likely detail-dependent. In
time-independent cases, the extremal
codim-1 surface volumes have dominant contributions from the near
boundary region (with cutoff $\epsilon$ and $d_i$ spatial dimensions),
giving the scaling\
$C(t)\propto {R^{d_i+1}\over G_{d_i+2}\,R}\, {V_{d_i}\over\epsilon^{d_i}}
\equiv N_{dof}\, V_{d_i}\Lambda_{_{UV}}^{d_i}$\,. This reflects the fact
that complexity scales with the number of degrees of freedom in the
dual field theory and with spatial volume in units of the UV
cutoff. Extensive further investigations of holographic complexity
including other proposals (such as complexity-equals-action,
complexity-equals-anything, path integral complexity), in particular
with relevance to cosmological contexts, appear in \eg\
\cite{Susskind:2015toa}-\cite{Aguilar-Gutierrez:2024rka}\ (see also the
review \cite{Chapman:2021jbh}).

It is fascinating to use such quantum information tools to probe
cosmological singularities which remain mysterious in many ways: we
will employ the ``complexity equals volume'' proposal (\ref{CV}) in
this regard. One might expect severe stringy and quantum effects to be
important here.  In the context of holography, one might imagine
certain classes of severe time-dependent deformations of the CFT to be
useful in shedding light on Big Bang/Crunch singularities.  A
prototypical example is the $AdS_5$ Kasner background, where the dual
Super Yang-Mills CFT can be regarded as subjected to severe time
dependent deformations (of the gauge coupling and the space on which
the CFT lives) \cite{Das:2006dz,Das:2006pw,Awad:2007fj,Awad:2008jf};
see also
\cite{Engelhardt:2014mea,Engelhardt:2015gta,Engelhardt:2015gla,Engelhardt:2016kqb} (and \cite{Craps:2006yb,Burgess:2011fa} for some reviews pertaining
to Big-Bang/Crunch singularities and string theory). These
are likely to be qualitatively different from bulk black holes
however, which being dual to thermal states might be regarded as
natural endpoints for generic time-dependent perturbations that would
thermalize on long timescales. There are indications that the dual
state to a Big-Bang/Crunch is quite non-generic. For instance, volume
complexity for $AdS$-Kasner singularities was found to become
vanishingly low in \cite{Barbon:2015ria} (see also
\cite{Caputa:2021pad}). This also appears consistent with the
investigations of classical and quantum codim-2 extremal surfaces and
holographic entanglement entropy in $AdS$ Kasner and other
singularities \cite{Manu:2020tty}, \cite{Goswami:2021ksw}: the
entangling surfaces are driven away from the near singularity region
(for spacelike singularities). These results suggest that the effective
number of qubits dual to the near singularity region is vanishingly
small, giving low complexity for the ``dual Kasner state'' independent
of the reference state, and low entanglement. The bulk singularity is
a Big-Bang/Crunch where spatial volumes, and thus the number of
degrees of freedom, become vanishingly small.  In some ways this might
naively contrast with colloquial thinking that a Big-Bang singularity
is a ``hot dense mess'' (\eg\ in FRW cosmologies) but perhaps reflects
the fact that these holographic singularities are low entropy
configurations.  Note also that in the eternal $AdS$ black hole, the
complexity extremal surfaces slice through the interior but stay well
away from the black hole singularity, approaching a limiting surface
for late times $t\ra\infty$\ (analogous to \cite{Hartman:2013qma}
for holographic entanglement entropy
\cite{Ryu:2006bv,Ryu:2006ef,HRT,Rangamani:2016dms}). This appears
to dovetail with the above, recalling the fact that the black hole
interior is a cosmology with a spacelike Big-Crunch singularity.


In the present paper we build on some of these investigations and
study holographic complexity and entanglement. Paraphrasing from
sec.~\ref{sec:lightliket(r)} and sec.~\ref{NCADSK}, we find that the
codim-1 complexity extremal surface anchored at some boundary time
slice $t$ some finite temporal distance from the singularity at $t=0$
begins as a spacelike surface near the boundary, bends away from the
singularity and approaches a lightlike trajectory in the $(t,r$)-plane
with $r$ the bulk holographic direction. As the anchoring time slice
$t$ becomes smaller, \ie\ going towards the singularity, the surface
transitions more quickly from spacelike towards the lightlike limit.
These features are depicted qualitatively in Fig.~\ref{figbbCcmpxtySurf}.
The lightlike part has vanishing volume so the complexity volume
functional becomes small as $t$ becomes small and eventually becomes
vanishingly small as $t\ra 0$ towards the singularity. This behaviour
is universal for various classes of Big-Bang/Crunch singularities,
viz. $AdS$ Kasner, hyperscaling violating Kasner and Lifshitz Kasner.
For $AdS$-Kasner and Lifshitz-Kasner backgrounds, complexity
decreases linearly with the anchoring boundary time slice $t$ as $t\ra
0$. For the hyperscaling violating Kasner backgrounds, the complexity
decrease is not linear with time $t$, reflecting the nontrivial
effective spatial dimension of the dual field theories. Overall our
results on complexity corroborate the discussions on complexity for
$AdS$ Kasner in \cite{Barbon:2015ria} but our analysis, especially
the extensive numerical studies, hopefully adds somewhat greater
detail. The complexity results for other cosmologies we discuss are
new but in accord with those for $AdS$ Kasner.

Our technical analysis for complexity has several parallels with that
in \cite{Manu:2020tty} of holographic entanglement entropy
\cite{Ryu:2006bv,Ryu:2006ef,HRT,Rangamani:2016dms}, in the
semiclassical region far from the singularity. Here we extend that
analysis numerically along the lines of volume complexity above for
entanglement as well. The codim-2 area functional has many technical
similarities with the codim-1 volume complexity functional so we find
similar results in the IR limit when the subregions are large and
essentially cover the whole space. The surface transitions from
spacelike near the boundary to lightlike in the interior. As the
anchoring time slice $t$ approaches the singularity, the lightlike
transition is more rapid. As $t\ra 0$, the entanglement entropy
becomes vanishingly small.

The vanishingly low complexity (and entanglement) for times approaching
the singularity reflect the fact that spatial volumes Crunch there, so
the effective number of degrees of freedom near the singularity is
vanishingly small.  From the point of view of constructing the dual
``Kasner'' state from some reference state, it appears that there are
simply vanishingly small numbers of effective qubits in the vicinity
of the singularity, independent of any reference initial state, thus
leading to low complexity.

The cosmological backgrounds are mostly isotropic, so the resulting
complexity volume functional can be recast in an effective
2-dimensional form, consistent with dimensional reduction of all the
boundary spatial dimensions. This results in a relatively simple
expression for complexity solely in terms of the variables describing
the effective 2-dim dilaton gravity theory arising under reduction,
\ie\ the 2-dim metric and the dilaton (which is the higher dimensional
transverse area). It may be interesting to interpret this effective
2-dimensional holographic complexity in terms of appropriate dual
effective 1-dim qubit models.

This paper is organized as follows. In sec.~\ref{HD-to-2D}, we argue
that volume complexity in these higher-dimensional theories can be
compactly recast as complexity in effective 2-dim theories that can be
regarded as arising by dimensional reduction. In
sec.~\ref{C-AdSK-sec}, we discuss holographic complexity of
$AdS$-Kasner: in sec.~\ref{C-AdS5K-subsec},
sec.~\ref{C-AdS4K-subsec}, and sec.~\ref{C-AdS7K-subsec}, we obtain
the solution of the equations of motion associated with complexity
surfaces for AdS$_{5,4,7}$-Kasner spacetimes which we then use to
obtain the holographic complexity of AdS$_{5,4,7}$-Kasner spacetimes
numerically in sec.~\ref{NCADSK}. Sec.~\ref{HSV-sec} discusses
holographic complexity in hyperscaling violating cosmologies,
focussing on $d_i=2,\ \theta=-{1\over 3}$ (sec.~\ref{D2}) and $d_i=4,
\ \theta=-1$ (sec.~\ref{D4}): we then numerically compute holographic
complexity in these cosmologies in sec.~\ref{NChvLif}. In
sec.~\ref{Lif-cosmo-sec}, we compute the holographic complexity of
isotropic Lifshitz Kasner cosmology. We review aspects of the
holographic entanglement studies in \cite{Manu:2020tty} in
sec.~\ref{EE-cosmologies}. We use this to discuss the behavior of
RT/HRT surfaces for $AdS$-Kasner spacetime in sec.~\ref{EE-AdSK-sec}
via sec.~\ref{EE-AdS5K-subsec} ($AdS_5$-Kasner) and
sec.~\ref{EE-AdS7K-subsec} ($AdS_7$-Kasner) and then compute
holographic entanglement entropy numerically in
sec.~\ref{NC-HEE-AdSKasner} of $AdS_{5,7}$-Kasner spacetimes.
Sec.~\ref{summary} contains a Discussion of our results alongwith
various comments and questions.

In App. \ref{sec:HolCosD}, we briefly review earlier studies on
holographic cosmologies and their 2-dim reduction and in App.
\ref{appendix-A}, we have listed the coefficients appearing in the
perturbative solution of AdS-Kasner and hyperscaling violating
cosmologies. In App. \ref{FA-EE-AdS5K}, we briefly discuss our
numerical methods applied to entanglement for finite subregions, with
the results vindicating expectations and thereby our overall analysis.

\section{\!\!\! Higher dim volume complexity\!  $\ra$ 2-dims: generalities}
\label{HD-to-2D}

The metric for an eternal $AdS_{d_i+2}$ Schwarzschild black hole (with
transverse space $d\sigma_{d_i}^2$) is
\be
ds^2 = {R^2\over r^2}\Big(-H(r)dt^2 + {dr^2\over H(r)} + d\sigma_{d_i}^2\Big)\,.
\ee
Then the complexity volume functional given by the volume of the
Einstein-Rosen bridge is
\be
C_D = {V_{d_i}\over G_{d_i+2} R} \int dr {R^{d_i+1}\over r^{d_i+1}}
\sqrt{{1\over H(r)} - H(r) t'(r)^2}\,.
\ee
Since the transverse space (that the codim-1 extremal surface wraps)
appears in a simple way in this expression, the complexity functional is
effectively 2-dimensional and can be recast explicitly in terms of
the complexity of an effective 2-dim dilaton gravity theory obtained
by dimensional reduction over the transverse space $d\sigma_{d_i}^2$.
This is quite general and applies for large families of backgrounds
that are ``mostly'' isotropic: the 2-dim dilaton gravity theory
for various purposes encapsulates the higher dimensional gravity
theory \cite{Narayan:2020pyj}. We will find this perspective useful
in what follows, where we study holographic backgrounds containing
cosmological singularities, particularly those studied in
\cite{Bhattacharya:2020qil}. A brief review of these studies and
holographic cosmologies appears in App.~\ref{sec:HolCosD}.

Consider the general ansatz for a $D=d_i+2$ dimensional gravity background
\bea\label{dimredAnsatz}
&& ds_D^2 = g_{\mu \nu}^{(2)} dx^\mu dx^\nu +\phi^{2/d_i}d\sigma_{d_i}^2
= \frac{e^f}{\phi^{(d_i-1)/d_i}}(-dt^2+dr^2)+\phi^{2/d_i}d\sigma_{d_i}^2\,,\nn\\
[1mm]
&& ds^2 = g_{\mu\nu} dx^\mu dx^\nu = e^f\,(-dt^2+dr^2)\,,\qquad 
g_{\mu\nu} = \phi^{(d_i-1)/d_i} g_{\mu \nu}^{(2)}\,.
\eea
Reviewing \cite{Narayan:2020pyj}, \cite{Bhattacharya:2020qil},
performing dimensional reduction over the transverse space
$d\sigma_{d_i}^2$ gives rise to a 2-dim dilaton gravity theory. With
the above parametrization, the higher dimensional transverse area is
the 2-dim dilaton $\phi$.
The Weyl transformation $g_{\mu \nu}=\phi^{(d_i-1)/d_i} g^{(2)}_{\mu \nu}$
absorbs the dilaton kinetic energy into the curvature $R^{(2)}$ and
the 2-dim action becomes
\be\label{2ddg-action}
S = {1\over 16\pi G_{2}} \int d^2x\sqrt{-g} \left(\phi {\cal R}
- U(\phi,\Psi) - {1\over 2}\,\phi (\del\Psi)^2\right) ,
\ee
with the dilaton potential $U(\phi,\Psi)$ potentially coupling
$\phi$ to another scalar $\Psi$ which is a minimal massless scalar
in the higher dimensional theory (see App.~\ref{sec:HolCosD}). The
dilaton factor in the $\Psi$
kinetic energy arises from the reduction to 2-dimensions. These models
with various kinds of dilaton potentials encapsulate large families
of nontrivial higher dimensional gravity theories with spacelike
Big-Bang/Crunch type cosmological singularities. In the vicinity of
the singularity, the 2-dim fields have power-law scaling behaviour
of the form (setting dimensionful scales to unity)
\be\label{2d-tr-exp}
\phi=t^kr^m\,,\qquad e^f=t^ar^b\,,\qquad e^\Psi=t^\al r^\beta\,,
\ee
and the forms of $e^f, \phi$ then translate to the higher dimensional
cosmological background profile containing the singularities. The 2-dim
formulation leads to various simplifications in the structure of these
backgrounds and reveals certain noteworthy features. In particular, the
severe time-dependence in the vicinity of the singularity implies that
the time derivative terms are dominant while other terms, in particular
pertaining to the dilaton potential, are irrelevant there. This then
reveals a ``universal'' subsector with $k=1$,
\be\label{universality}
\phi\sim t\,,\qquad e^f\sim t^a\,,\qquad e^\Psi\sim t^\al\,;\qquad\quad
a={\al^2\over 2}\,.
\ee
A prototypical example is $AdS$ Kasner \cite{Das:2006dz} and its
reduction to 2-dimensions,
\bea\label{AdSDK-2d-0}
&&  ds^2 = {R^2\over r^2} (-dt^2 + dr^2) + {t^{2/d_i}\,R^2\over r^2} dx_i^2\,,
\qquad e^\Psi = t^{\sqrt{{2(d_i-1)/d_i}}}\,,\qquad
\Lambda=-{d_i(d_i+1)\over 2R^2}\,,\quad \nn\\ [1mm]
\ra && \qquad\ \phi={t\,R^{d_i}\over r^{d_i}}\,,\qquad
ds^2={t^{(d_i-1)/d_i}\,R^{d_i+1}\over r^{d_i+1}}(-dt^2+dr^2)\,,\qquad
U=2\Lambda\phi^{1/d_i}\,.
\eea
The isotropic restriction from general $AdS$ Kasner (\ref{adsKasner})
alongwith the Kasner exponent relation
$\sum_ip_i=1$ implies a single Kasner exponent $p={1\over d_i}$.
$R$ is the $AdS$ length scale. We are suppressing an implicit Kasner
scale $t_K$: \eg\ $t^{2p}\ra (t/t_K)^{2p}$. We will reinstate this as
required. There are several more general families of such backgrounds
with Big-Bang/Crunch singularities, including nonrelativistic ones
such as hyperscaling violating (conformally $AdS$) theories, and 
those with nontrivial Lifshitz scaling, as we will discuss in
what follows, and summarized in the Table \ref{table-exponents}.
\begin{table}[h]
\begin{center}
\begin{tabular}{ |c|c|c|c|c|} 
 \hline
{\bf Cosmologies} &  $k$ & $m$ & $a={\al^2\over 2}$ &  $b$  \\
 \hline 
$AdS$ Kasner cosmology\ & 1 & $-d_i$ & $\frac{d_i-1}{d_i}$  & $-(d_i+1)$ \\ 
 \hline
 Hv cosmology ($z=1,\ \theta\neq 0$)\ & 1 & $-(d_i-\theta)$ & $\left(\sqrt{\frac{d_i-\theta -1}{d_i-\theta }}-\sqrt{\frac{(-\theta) }{d_i (d_i-\theta
   )}}\right)^2$ & $-\frac{(d_i-\theta)(1+d_i)}{d_i}$ \\ 
 \hline 
Lif cosmology ($z=d_i,\ \theta=0$)\ & 1 & $-1$ & ${d_i-1\over d_i}$ & $-3+{1\over d_i}$ \\ 
 \hline
\end{tabular}
\end{center}
\caption{Exponents for 2-dim cosmologies.}
\label{table-exponents}
\end{table}

The time dependence in these backgrounds does not switch off
asymptotically so that simple interpretations in terms of deformations
of some vacuum state appear difficult: instead these are probably best
regarded as dual to some nontrivial nongeneric state in the dual field
theory. This is consistent with the expectation that generic severe
time-dependent CFT deformations will thermalize and thus be dual to
black hole formation in the bulk. Further discussions on this
perspective appear throughout the paper (building on
\cite{Das:2006dz,Das:2006pw,Awad:2007fj,Awad:2008jf}, and
\cite{Manu:2020tty}, \cite{Goswami:2021ksw}).

We now come to complexity. We mostly consider the
transverse space to be planar, so $d\sigma_{d_i}^2 = \sum_idx_i^2$.
Then, in terms of 2-dim variables (\ref{dimredAnsatz}) the complexity
volume functional becomes
\bea\label{Volume-CES}
C &=& {1\over G_{d_i+2} R} \int \prod_{j=1}^{d_i}\left(\phi^{1/d_i}dx_j\right) \sqrt{\frac{e^f}{\phi^{(d_i-1)/d_i}}(-dt^2+dr^2)} \nn\\   
&=& {V_{d_i}\over G_{d_i+2} R} \int_\epsilon dr\ \phi^{\frac{(d_i+1)}{2 d_i}}\,
e^{f/2}\, \sqrt{1-t'(r)^2}\
=\ {1\over G_{2} R} \int_\epsilon dr\ \phi^{\frac{(d_i+1)}{2 d_i}}\,
e^{f/2}\, \sqrt{1-t'(r)^2}\,,\qquad
\eea
with $G_2= {G_{d_i+2}\over V_{d_i}}$ the 2-dimensional Newton constant
after reduction, and $r=\epsilon$ the holographic boundary. The higher
dimensional curvature scale (\eg\ $R$ in (\ref{AdSDK-2d-0})) continues
as the 2-dim curvature scale.
Also, $t'\equiv {dt\over dr}$ is the $r$-derivative of the time coordinate
as a function $t(r)$ of the holographic radial coordinate.

The last expression in (\ref{Volume-CES}) above can be interpreted as
the complexity volume functional in the 2-dim dilaton gravity theory
intrinsically. It would then be interesting to ask for dual
1-dimensional effective qubit models whose complexity can be understood
as this.

Sticking in the power-law ansatze (\ref{2d-tr-exp}) above, we obtain
\be\label{Complexity-behavior-NHM}
C = \frac{V_{d_i}}{G_{d_i+2} R} \int_\epsilon dr\,\ t(r)^{\left(k\left(\frac{(d_i+1)}{2 d_i}\right)+\frac{a}{2}\right)}\ r^{\left(m\left(\frac{(d_i+1)}{2 d_i}\right)+\frac{b}{2}\right)}\ \sqrt{1-t'(r)^2}\ 
\equiv\
\frac{V_{d_i}}{G_{d_i+2} R} \int_\epsilon dr\, {\cal L}\,,
\ee
with ${\cal L}\equiv {\cal L}\left(r,t(r),t'(r)\right)$ the effective
Lagrangian. Extremizing for the complexity surface $t(r)$ leads to the
Euler-Lagrange equation\
$\frac{d}{dr}\big(\frac{\partial {\cal L}}{\partial t'(r)} \big)-\frac{\partial {\cal L}}{\partial t(r)}=0$\,.\ \
Simplifying this gives the equation of motion for the complexity
surface $t(r)$ in (\ref{Complexity-behavior-NHM})
\bea\label{EOM-t[r]}
2 d_i r\, t\, t'' + r\, (a d_i+d_i+1)\, \left(1-(t')^2\right)
+ (b d_i+d_i m+m)\,t\,t' \left(1-(t')^2\right) = 0\,,
\eea
abbreviating notation with
$t\equiv t(r),\ t'\equiv {dt\over dr}\,,\ t''\equiv {d^2t\over dr^2}$,
and we have used the universality result $k=1$ in (\ref{universality}).

Now we solve equation (\ref{EOM-t[r]}) perturbatively and numerically
for $AdS$ Kasner, hyperscaling violating and Lifshitz cosmologies, and
thereby compute holographic complexity for these cosmologies
in sections \ref{C-AdSK-sec} and \ref{HSV-sec}.

\medskip

\noindent {\bf Methodology}:\ We outline our techniques and methods here:
\begin{enumerate}
\item For a given background, first, we solve (\ref{EOM-t[r]})
  semiclassically in perturbation theory using an ansatz of the form
$t(r)=\sum_{n \in \mathbb{Z}_+}c_n r^n$ for the complexity surfaces
$t(r)$, as functions of the radial coordinate $r$ for various anchoring
time slices $t_0$ which define boundary conditions at $r=0$. The
perturbative solutions are valid only in a certain $r$-regime,
\ie\ upto a cut-off $r_\Lambda$\ (roughly $r_\Lambda \lesssim t_0$).
Thus these cannot encapsulate the entire bulk geometry.
\item To overcome this and obtain a global picture of the bulk, we 
solve (\ref{EOM-t[r]}) numerically (in Mathematica). For this purpose,
we need two initial conditions which we extract from the perturbative
solutions for $t(r)$ above and their derivatives $t'(r)$, setting
the boundary value $r=\epsilon=10^{-2}$ and $t_0$ as the numerical
value of a specific anchoring time slice (with all other dimensionful
scales set to unity). This allows us to obtain numerical solutions
for the complexity surfaces, which then reveal nontrivial bulk
features such as lightlike limits and the transition thereto, from
spacelike regimes near the boundary. This then allows us to numerically
evaluate holographic volume complexity and plot it against $t_0$
for various backgrounds.
\item We then employ similar algorithms broadly for holographic
entanglement entropy.
\item Some numerical issues persist for certain backgrounds, as we
state in what follows, and we suppress detailed discussions in
these cases.
\end{enumerate}

\section{Complexity:\ $AdS$ Kasner}\label{C-AdSK-sec}

The isotropic $AdS_{d_i+2}$ Kasner spacetime (\ref{AdSDK-2d-0}) in
the form of the reduction ansatz (\ref{dimredAnsatz}) alongwith its
2-dim exponents (\ref{2d-tr-exp}), is:
\bea\label{AdSDK-2d}
&&  ds^2 = {R^2\over r^2} (-dt^2 + dr^2) + {t^{2/d_i}\,R^2\over r^2} dx_i^2\,,
\qquad e^\Psi = t^{\sqrt{{2(d_i-1)/d_i}}}\,,\qquad
\Lambda=-{d_i(d_i+1)\over 2R^2}\,,\qquad \nn\\ [1mm]
&& \quad \phi={t\,R^{d_i}\over r^{d_i}}\,,\qquad
ds^2={t^{(d_i-1)/d_i}\,R^{d_i+1}\over r^{d_i+1}}(-dt^2+dr^2)\,,\qquad
U=2\Lambda\phi^{1/d_i}\,,\nn\\
&& \qquad k=1\,, \quad\ \ m=-d_i\,, \quad\ \ a=\frac{d_i-1}{d_i}\,,
\quad\ \ b=-(d_i+1)\,, \quad\ \ \al = \sqrt{{2(d_i-1)\over d_i}}\ .\quad
\eea
The single Kasner exponent $p={1\over d_i}$ arises due to the
isotropic restriction in (\ref{adsKasner}).\ $R$ is the $AdS$ length
scale. We are suppressing an implicit Kasner scale $t_K$: \eg\
$t^{2p}\ra (t/t_K)^{2p}$.

Then the $t(r)$ extremization equation (\ref{EOM-t[r]}) becomes
\begin{eqnarray}
\label{EOM-t[r]-AdS-Kasner}
& &  r\,t(r)\, t''(r) - ({d_i}+1)\,t(r)\,t'(r)\,\left(1- t'(r)^2 \right) 
+ r\,\left(1-t'(r)^2\right) = 0\,.
\end{eqnarray}
We discuss the solution of (\ref{EOM-t[r]-AdS-Kasner}) for
$AdS_{5,4,7}$-Kasner spacetimes in sec.~\ref{C-AdS5K-subsec},
\ref{C-AdS4K-subsec}, and \ref{C-AdS7K-subsec}.

\subsection{AdS$_5$-Kasner spacetime}\label{C-AdS5K-subsec}

For $AdS_5$-Kasner spacetime, we have $d_i=3$: then
(\ref{EOM-t[r]-AdS-Kasner}) simplifies to
\begin{eqnarray}
\label{EOM-t[r]-AdS5-Kasner}
& &  r\,t(r)\, t''(r) - 4 t(r)\, t'(r)\,\left(1-t'(r)^2\right)
+ r\,\left(1 - t'(r)^2\right) = 0\,.
\end{eqnarray}
First we note that with $t', t''=0$, the equation above is not
satisfied except for $r\sim 0$, so that $t(r)=const$ is not a solution:
the surface always bends in the time direction due to the time
dependence of the background.\ When the complexity
surface $t(r)$ has weak $r$-dependence, \ie\ it is almost constant
with $t(r)\sim t_0$, we can analyze the above equation in perturbation
theory in $r$, by considering the following ansatz for $t(r)$:
\begin{eqnarray}
\label{ansatz-t[r]}
t(r)=t_0+ \sum_{n \in \mathbb{Z}_+} c_n r^n\,.
\end{eqnarray}
Inputting this ansatz (\ref{ansatz-t[r]}) in (\ref{EOM-t[r]-AdS5-Kasner})
and solving for the coefficients iteratively, we found the 
solution up to ${\cal O}(r^{30})$. Up to $O(r^4)$, this is 
\begin{eqnarray}
\label{soln-t[r]-30}
& & \hskip -0.2in
t(r)=t_0+\frac{r^2}{6
   t_0}-\frac{7 r^4}{216 t_0^3}\ .
\end{eqnarray}
We have truncated the solution (\ref{soln-t[r]-30}) up to $O(r^4)$ here
for brevity of the series expansion (this is $O(r^3)$ in $t'(r)$ in
(\ref{t'[r]})). A more detailed iterative version of (\ref{soln-t[r]-30}) 
up to $O(r^{30})$ appears in (\ref{soln-t[r]-30-app}).
Likewise, truncated solutions are displayed elsewhere in the paper, \eg\
(\ref{soln-t[r]-ii}), (\ref{soln-t(r)-AdS7K}), (\ref{t[r]-general-soln-NCB-IR-theta=-1by3}), (\ref{t[r]-soln-NCB-IR-theta=-1by3}), (\ref{t[r]-general-soln-NCB-IR-theta=-1}), (\ref{t[r]-soln-NCB-IR-theta=-1}) and (\ref{lifshitz_metric_z=2,d=2, theta=0_pertubative_sol_t}).
The numerical analysis described in what follows is based on the more
detailed expansion up to $O(r^{30})$. However we find that the numerical
plots do not change much with the truncated solutions such as
(\ref{soln-t[r]-30}): the qualitative features of the extremal surfaces
are the same. So we will continue to display the truncated solutions
alone in the rest of the paper for compactness.

\begin{figure}
\begin{subfigure}
  \centering
  \includegraphics[width=.5\linewidth]{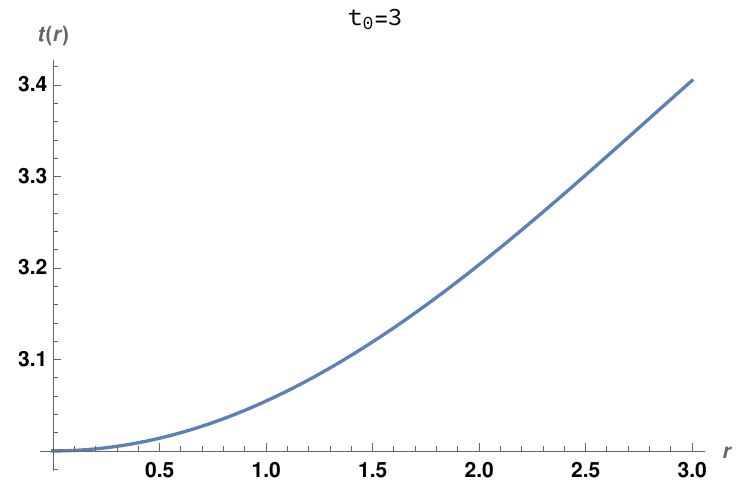}
\end{subfigure}%
\begin{subfigure}
  \centering
  \includegraphics[width=.5\linewidth]{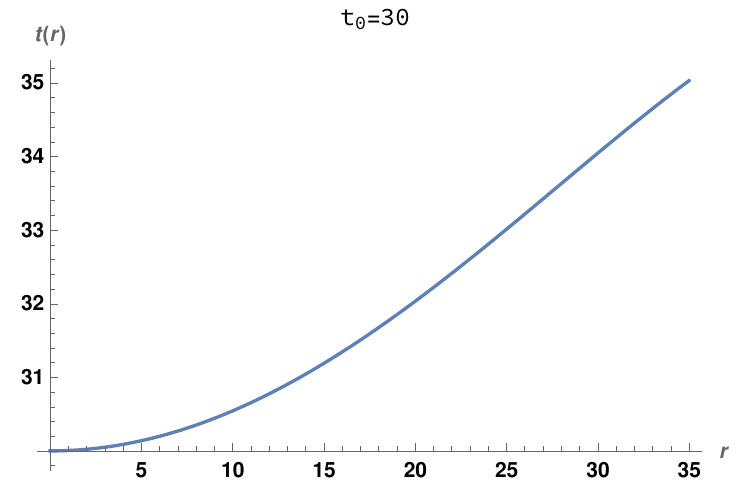}
\end{subfigure}
\begin{subfigure}
  \centering
  \includegraphics[width=.5\linewidth]{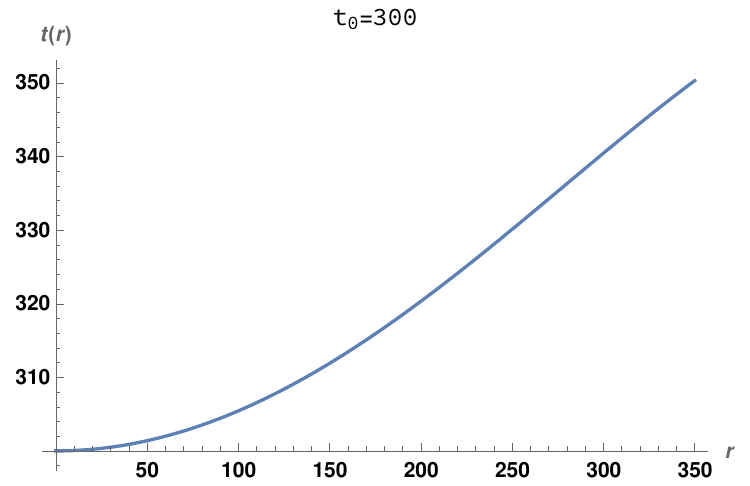}
\end{subfigure}
\begin{subfigure}
  \centering
  \includegraphics[width=.5\linewidth]{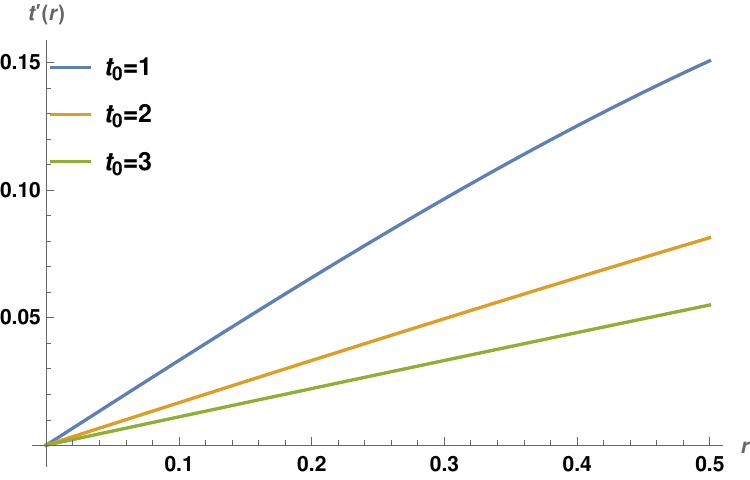}
\end{subfigure}%
\begin{subfigure}
  \centering
  \includegraphics[width=.5\linewidth]{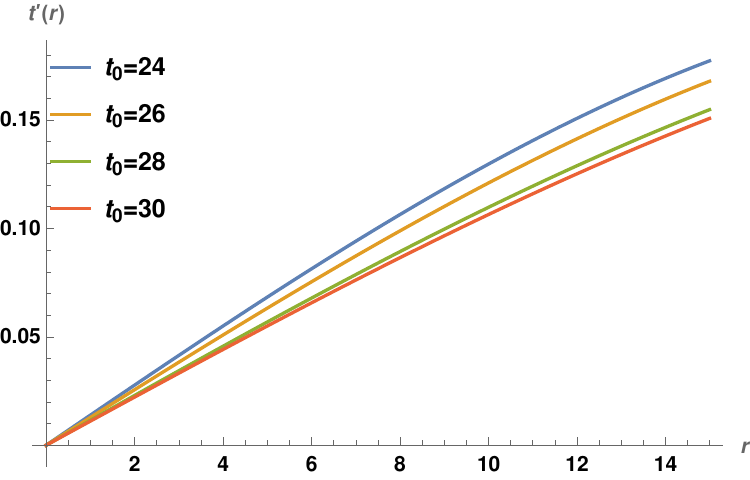}
\end{subfigure}
\begin{subfigure}
  \centering
  \includegraphics[width=.5\linewidth]{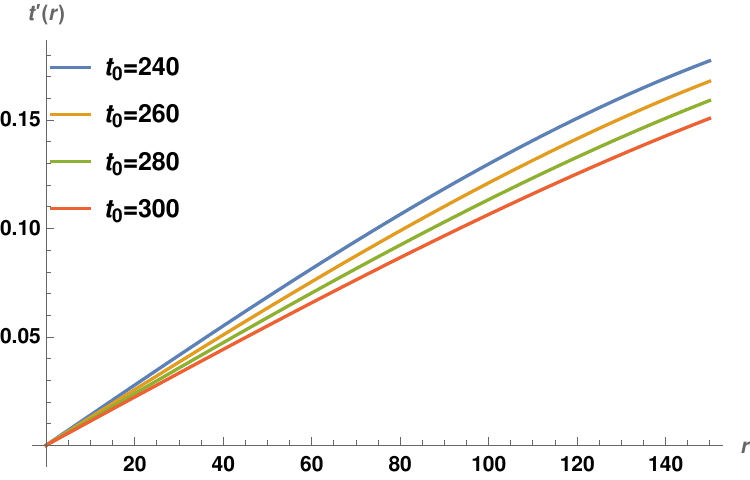}
\end{subfigure}
\caption{Variation with $r$ of the semiclassical (perturbative)
  complexity surfaces $t(r)$ and their derivatives $t'(r)$ for various
  values of the anchoring time slice $t_0$.}
\label{t[r]-plots}
\end{figure}

The solution $t(r)$ in (\ref{soln-t[r]-30-app}) and its derivative $t'(r)$
are plotted in Fig.~\ref{t[r]-plots} for various $t_0$-values\footnote{We
  obtain similar qualitative behaviour for the variation of complexity
  surfaces and their derivatives in other backgrounds, so we will not
  display them, in favour of the plots of numerical solutions which
  are more instructive.}.
From Fig.~\ref{t[r]-plots}, we see that the complexity surface varies
approximately linearly with $r$ in the regime roughly $r \sim t_0$ and
$t'(r)$ reaches its maximum value about $t'(r)\lesssim 0.2$, which
vindicates the mild bending of the surface $t(r)$ in the radial
direction. Of course this perturbative solution has clear limitations,
expressed as it is here by a finite power series. However it is of
great value to display the behaviour of the complexity surface near
the boundary $r=0$.

We expect that when the anchoring time slice $t_0$ is far from the
singularity at $t=0$, the above perturbative solution is reasonable,
at least in the neighbourhood of the boundary. A very similar analysis
was carried out in \cite{Manu:2020tty} to reveal the RT/HRT surface in
the semiclassical regime far from the singularity bends away from the
singularity. We will study this numerically later revealing more
information.

A further check of the above series solution is that in the
semiclassical limit, when we ignore the higher order terms in $t'(r)$
(\ie\ $t'(r)\ll 1$), then (\ref{EOM-t[r]-AdS5-Kasner}) reduces to
\begin{eqnarray}
\label{EOM-t[r]-AdS5-Kasner-i}
& & t(r) \left(r t''(r)-4t'(r)\right)+r=0\,.
\end{eqnarray}
Solving (\ref{EOM-t[r]-AdS5-Kasner-i}) with the ansatz
(\ref{ansatz-t[r]}), we obtain up to ${\cal O}(r^{4})$:
{
\begin{eqnarray}
\label{soln-t[r]-ii}
& & \hskip -0.2in
t(r)=t_0+\frac{r^2}{6 t_0}-\frac{r^4}{24 t_0^3}\,.
\end{eqnarray}
}
Plotting (\ref{soln-t[r]-ii}) and its $r$-derivative reveals that 
the behaviour of the complexity surface $t(r)$ and its derivative is
qualitatively similar to that in Fig.~\ref{t[r]-plots}. This vindicates
the fact that $t'(r)$ is indeed small in this regime.

{\bf Holographic complexity of AdS$_5$-Kasner spacetime}:\
The holographic volume complexity (\ref{Complexity-behavior-NHM}) for
the $AdS_5$-Kasner spacetime (\ref{AdSDK-2d}) with $d_i=3$ simplifies to
\begin{eqnarray}
\label{Complexity-behavior-NHM-AdS5-Kasner}
C=\frac{V_{3}{R^{3}}}{G_{5}} \int_\epsilon dr \left(\frac{t(r) \sqrt{\left(1-t'(r)^2\right)}}{r^4}\right).
\end{eqnarray}
Now we compute this for the solutions (\ref{soln-t[r]-ii}) and (\ref{soln-t[r]-30}).

The semiclassical solution (\ref{soln-t[r]-ii}) was obtained with
$t'(r)\ll 1$ so we can approximate the complexity functional
(\ref{Complexity-behavior-NHM-AdS5-Kasner}) as:
\begin{eqnarray}
\label{Complexity-behavior-NHM-AdS5-Kasner-t'(r)<1}
C \sim \frac{V_{3}{R^{3}}}{G_{5}} \int_\epsilon^{r_{\Lambda}} dr \left(\frac{t(r) {\left(1-\frac{t'(r)^2}{2}\right)}}{r^4}\right).
\end{eqnarray}
We have inserted a cut-off $r_\Lambda$ in (\ref{Complexity-behavior-NHM-AdS5-Kasner-t'(r)<1}) because the perturbative solution is only valid upto some
$r_\Lambda\lesssim t_0$, and so this only covers part of the full complexity
surface. Beyond this we require additional analysis, which we carry
out numerically later.

Substituting the semiclassical solution $t(r)$ from (\ref{soln-t[r]-ii})
into (\ref{Complexity-behavior-NHM-AdS5-Kasner-t'(r)<1}) and integrating
gives complexity as (writing only terms up to next-to-leading order in
$t_0$ for simplicity)
\begin{eqnarray}
\label{C-30}
& & \hskip -0.3in {C}(t_0,r_\Lambda,\epsilon) \approx \frac{V_{3}{R^{3}}}{G_{5}}\Biggl[t_0 \left(\frac{0.3}{\epsilon ^3}-\frac{0.3}{r_\Lambda^3}\right)+\frac{1}{t_0}\left(\frac{0.1}{\epsilon
   }-\frac{0.1}{r_\Lambda}\right)+O\left(\left(\frac{1}{t_0}\right)^3\right)\Biggr].
\end{eqnarray}
For the more general solution (\ref{soln-t[r]-30}) obtained retaining all
the nonlinear terms in (\ref{EOM-t[r]-AdS5-Kasner}), we find $t'(r)$ as:
\begin{eqnarray}
\label{t'[r]}
& & \hskip -0.3in t'(r) \approx \frac{0.3
   r}{t_0} -\frac{0.1 r^3}{t_0^3}.
\end{eqnarray}
Thus we see that $t'(r)\ll 1$ provided $r \lesssim t_0$. In this
approximation, we can evaluate complexity as
(\ref{Complexity-behavior-NHM-AdS5-Kasner-t'(r)<1}) with the solution
(\ref{soln-t[r]-30}). Then holographic complexity is the same as
(\ref{C-30}) upto next-to-leading-order in $t_0$.

Going beyond perturbation theory fascinatingly shows that the
complexity surface becomes lightlike in the interior, as first noted
in \cite{Barbon:2015ria}.  This can be seen right away by noting that
(\ref{EOM-t[r]-AdS5-Kasner}) is in fact satisfied identically when
$t'(r)=1$ and $t''(r)=0$, \ie\ with $t(r)\sim r$ being lightlike
independent of the anchoring time slice $t_0$.

Towards identifying this in the $AdS_5$-Kasner spacetime, consider the
following ansatz for $t(r)$ around the lightlike limit with $f(r)$ a
small deviation:
\begin{eqnarray}
\label{t(r)-LL}
& & t(r)=t_0+r+f(r)\,.
\end{eqnarray}
This ansatz simplifies (\ref{EOM-t[r]-AdS5-Kasner}) to
\begin{eqnarray}
\label{EOM-f[r]}
& & \hskip -0.2in -r \left(f'(r)+1\right)^2+(f(r)+r+t_{0}) \left(r f''(r)+4 f'(r) \left(f'(r)+1\right)
\left(f'(r)+2\right)\right)+r=0\,.\qquad
\end{eqnarray}
Linearizing the above equation, \ie\ ignoring higher order terms in
$f'(r)$, gives
\begin{eqnarray}
\label{EOM-f[r]-linear}
& & r^2 f''(r)+r t_{0} f''(r)+8 t_{0} f'(r)+6 r f'(r)=0\,,
\end{eqnarray}
which can be solved as
\begin{eqnarray}
\label{soln-f[r]}
& & f(r)=c_1 \left(-\frac{t_{0}^2}{7 r^7}-\frac{t_{0}}{3 r^6}-\frac{1}{5 r^5}\right)+c_2\,,
\end{eqnarray}
where $c_1, c_2$ are constants. This gives the solution for $t(r)$ as
\begin{eqnarray}
\label{t(r)-LL-i}
& & t(r)=t_0+r+c_1 \left(-\frac{t_{0}^2}{7 r^7}-\frac{t_{0}}{3 r^6}-\frac{1}{5 r^5}\right)+c_2\,.
\end{eqnarray}
The above solution is not well-behaved when extrapolated all the way
to the boundary $r=0$ but it indicates the existence of the neighbourhood
of a lightlike surface. We now look for the lightlike solution numerically.

\subsubsection{Lightlike limits of complexity surfaces, numerically}
\label{sec:lightliket(r)}

Now we solve equation (\ref{EOM-t[r]-AdS5-Kasner}) numerically. Since
this is a second-order nonlinear differential equation, we need two
initial conditions for a numerical solution. One trivial initial
condition is $t(r=0)=t_0$, leaving the question of the initial condition
for $t'(r=0)$. Since we have solved (\ref{EOM-t[r]-AdS5-Kasner})
perturbatively obtaining (\ref{soln-t[r]-30}), we can obtain the
initial condition $t'(r=0)$ by evaluating the $r$-derivative thereof.
We regulate the holographic boundary $r=\epsilon\sim 0$ by choosing
$\epsilon=10^{-2}$ as the boundary point. For a specific slice $t_0$, we can
obtain initial conditions $t(r=0.01)$ and $t'(r=0.01)$ by substituting
$r=0.01$ and the value of $t_0$ in the solution (\ref{soln-t[r]-30}) and
its $r$-derivative\footnote{We have used this method in obtaining the
  numerical solution of the equation of motion associated with
  complexity/entanglement surfaces throughout the paper for different
  backgrounds. Therefore, we will not repeat this again: we will
  simply quote the results for different backgrounds.}.
The numerical solutions have been carried out in Mathematica, finetuning
the accuracy to required extent for the initial conditions, in particular
setting WorkingPrecision to MachinePrecision and PrecisionGoal to
Infinity (in using NDSolve):\ without these the results we obtained
were not adequately clean, and it took some attempts (over a long
while!) to tweak our numerics\ (the Mathematica files are available
upon request). Some of these results have been cross-checked and
corroborated via Python codes as well. Presumably the numerics can
be improved further.

\begin{figure}[h]
\begin{subfigure}
  \centering
  \includegraphics[width=.45\linewidth]{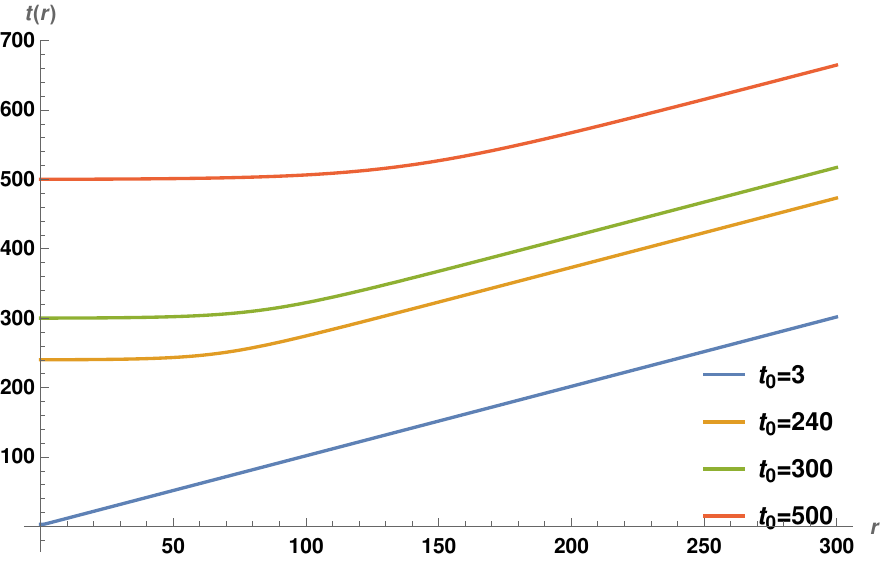}
\end{subfigure}
\begin{subfigure}
  \centering
  \includegraphics[width=.45\linewidth]{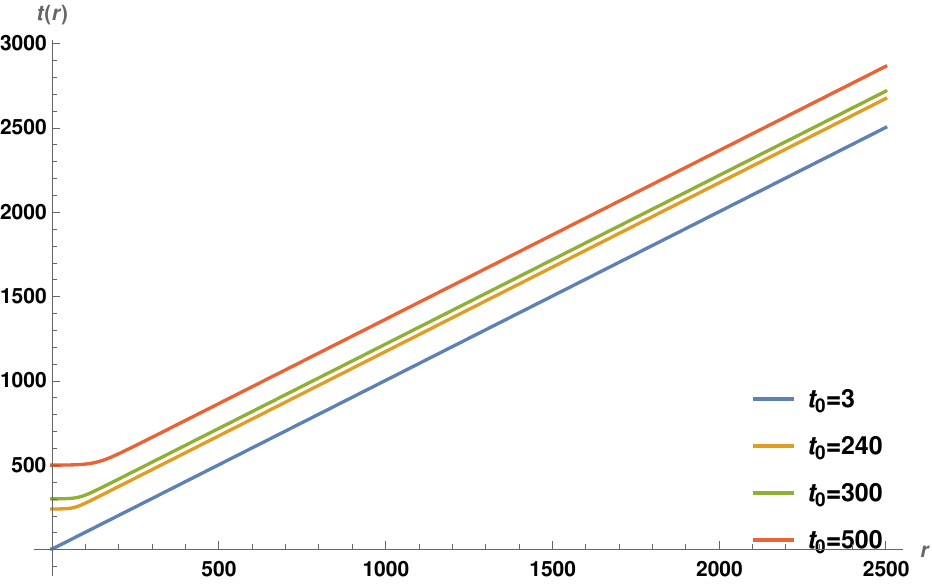}
\end{subfigure}
\caption{Numerical plots of the complexity surface versus $r$ in AdS$_5$-Kasner spacetime for different slices of $t_0$. In the right Figure, we have extended the range of radial coordinate.}
\label{LL-i-t[r]}
\end{figure}

\begin{figure}
\begin{subfigure}
  \centering
  \includegraphics[width=.5\linewidth]{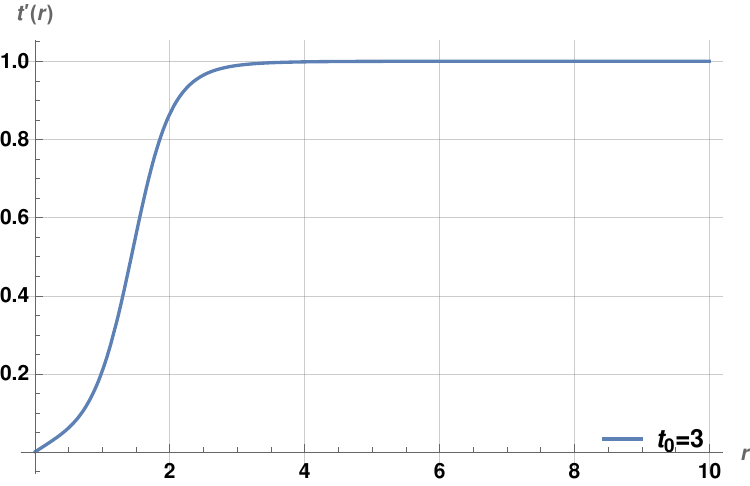}
\end{subfigure}
\begin{subfigure}
  \centering
  \includegraphics[width=.5\linewidth]{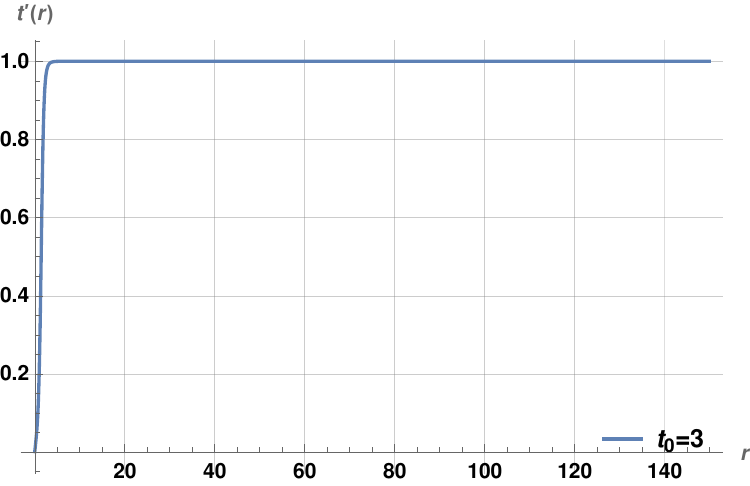}
\end{subfigure}
\begin{subfigure}
  \centering
  \includegraphics[width=.5\linewidth]{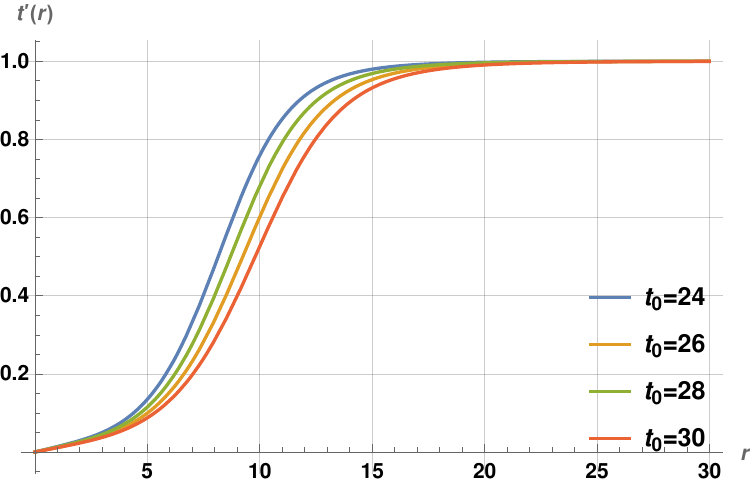}
\end{subfigure}
\begin{subfigure}
  \centering
  \includegraphics[width=.5\linewidth]{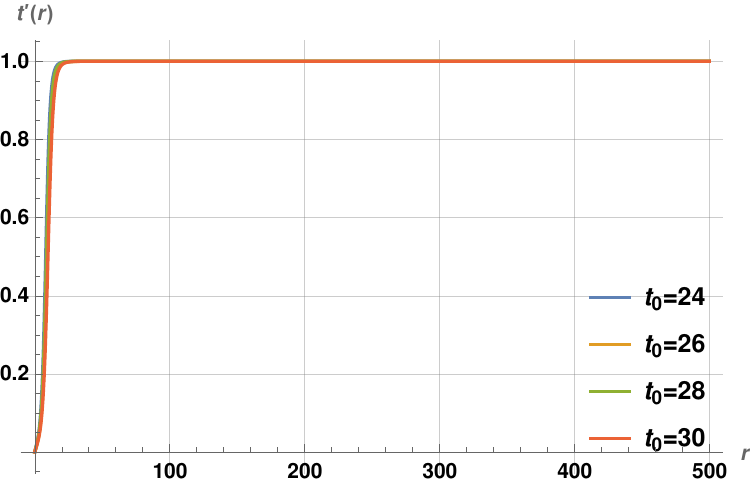}
\end{subfigure}
\begin{subfigure}
  \centering
  \includegraphics[width=.5\linewidth]{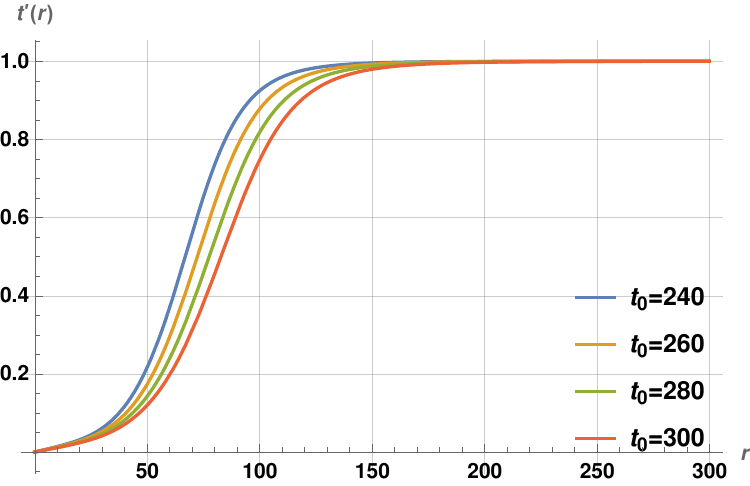}
\end{subfigure}
\begin{subfigure}
  \centering
  \includegraphics[width=.5\linewidth]{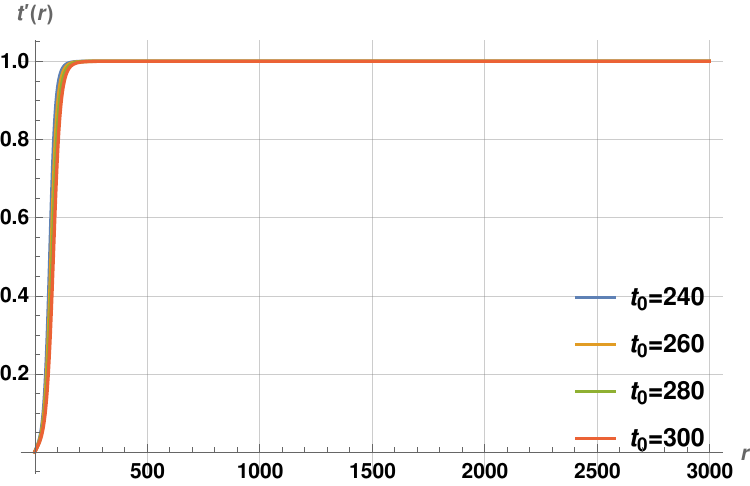}
\end{subfigure}
\caption{Numerical plots of $t'(r)$ versus $r$ in $AdS_5$-Kasner spacetime for different $t_0$ slices. In the right Figures, we have extended the range of the radial coordinate.}
\label{LL-i}
\end{figure}

The numerical solutions of (\ref{EOM-t[r]-AdS5-Kasner}) for the
complexity surface $t(r)$ and
their derivatives $t'(r)$ for different $t_0$ slices are plotted in
Fig.~\ref{LL-i-t[r]} and Fig.~\ref{LL-i} respectively. Some striking
points to note are:
\begin{itemize}
\item In Fig.~\ref{LL-i-t[r]}, we remind the reader that $t(r)$
  corresponds to $|t(r)|$ so the singularity is at $t=0$ (the horizontal
  axis at the bottom). Thus all complexity surfaces bend away from the
  neighbourhood of the singularity, which correlates with $t'(r)>0$.
\item From Fig. \ref{LL-i-t[r]}, the complexity surfaces become lightlike
after a certain value of $r$ for any anchoring time slice $t_0$.
\item The surfaces with lower $t_0$ (\ie\ closer to the singularity)
  become lightlike earlier (at smaller $r$) than those with larger
  $t_0$. This is also vindicated in Fig.~\ref{LL-i}, where we have
  numerically plotted $t'(r)$ with $r$.  All the complexity surfaces
  $t(r)$ approach $t'(r)=1$ eventually, \ie\ a lightlike regime.
\item The lightlike regime $t'(r)=1$ implies vanishing holographic
complexity here from the $\sqrt{1-t'(r)^2}$ factor in 
(\ref{Complexity-behavior-NHM-AdS5-Kasner}). Thus, numerically we
see that complexity picks up finite contributions only from the
near-boundary spacelike part of the complexity surfaces, beyond which
it has negligible value where the complexity surfaces are lightlike.
\item The above two points imply that as the anchoring time slice
  approaches the singularity location $t_0\ra 0$, the complexity surface
  is almost entirely lightlike: thus as $t_0\ra 0$ the holographic
  volume complexity becomes vanishingly small. We verify this later
  by numerical evaluation of the volume complexity integral in sec. \ref{NCADSK}.
\item These numerical plots and this analysis only makes sense for
  $t_0$ not strictly vanishing (\eg\ we require $t_0\gtrsim\epsilon$).
  In close proximity to the singularity, the semiclassical gravity
  framework here and our analysis breaks down.
\end{itemize}

In Figs.~\ref{LL-i-t[r]} and \ref{LL-i}, we have shown the behaviour
of $t(r)$ and $t'(r)$ with $r$ for both limited range (left side
plots) and extended range of the radial coordinate in these Figures
(right side plots). We obtain similar behavior for other cases
later. So we will not show the counterparts of the right side plots
(extended $r$-range) in Figs.~\ref{LL-i-t[r]}-\ref{LL-i} in order 
to display our results succinctly in subsequent data.

\subsection{Holographic complexity of $AdS_4$-Kasner spacetime}
\label{C-AdS4K-subsec}
\begin{figure}[h]
\begin{subfigure}
  \centering
  \includegraphics[width=.5\linewidth]{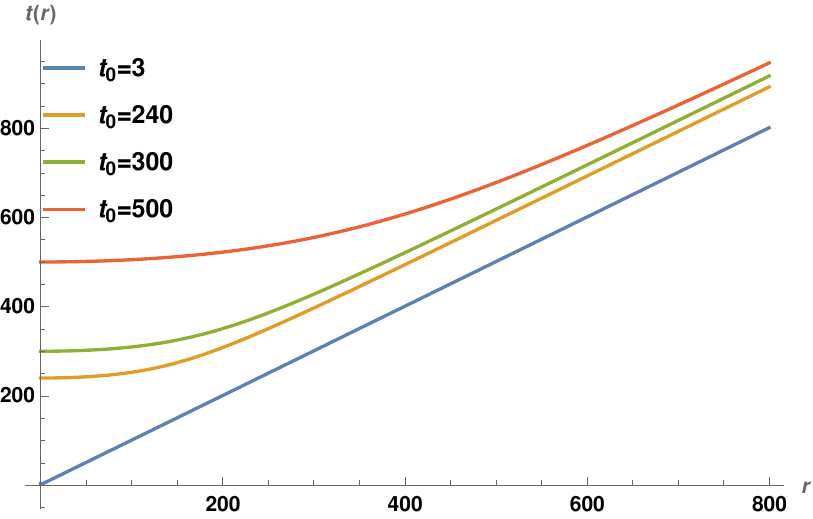}
\end{subfigure}
\begin{subfigure}
  \centering
  \includegraphics[width=.5\linewidth]{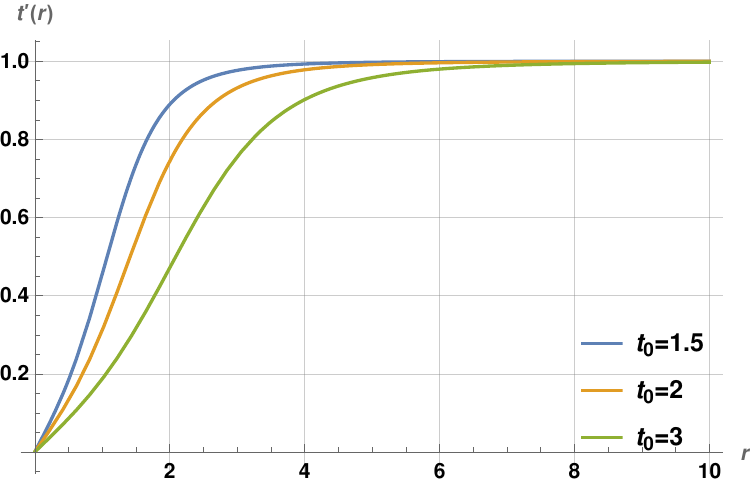}
\end{subfigure}
\begin{subfigure}
  \centering
  \includegraphics[width=.5\linewidth]{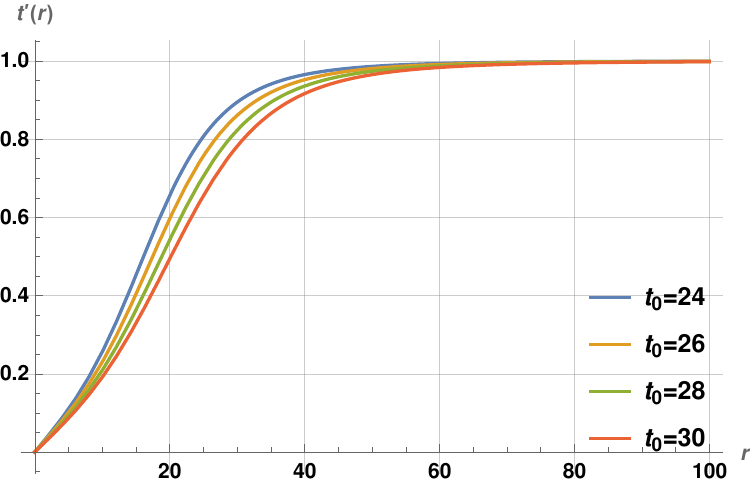}
\end{subfigure}
\begin{subfigure}
  \centering
  \includegraphics[width=.5\linewidth]{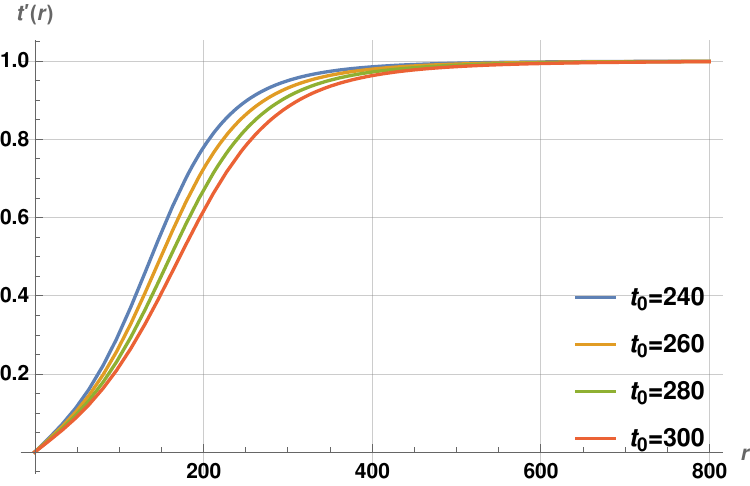}
\end{subfigure}
\caption{Plots of $t(r)$ with $r$ and $t'(r)$ with $r$ in
  $AdS_4$-Kasner spacetime for various $t_0$ slices. }
\label{LL-i-AdS4K-Num}
\end{figure}
For the $AdS_4$-Kasner spacetime with $d_i=2$, the equation of motion
(\ref{EOM-t[r]-AdS-Kasner}) for the complexity surface $t(r)$ becomes
\begin{eqnarray}
\label{EOM-AdS4K}
r\,t(r)\,t''(r)
-3t(r)\,t'(r) \left(1-t'(r)^2\right) + r \left(1-t'(r)^2\right) = 0\,.
\end{eqnarray}
{\bf Numerical results}:\ 
The perturbative solution of (\ref{EOM-AdS4K}) is obtained only up to
$O(r^2)$ unfortunately, \ie\ $t(r)=t_0+\frac{r^2}{4 t_0}$\,, beyond
which the numerics appear problematic.  However, we find that this
$O(r^2)$ perturbative solution is adequate in extracting the initial
conditions for numerical solutions.

Using these initial conditions, the numerical solution of
(\ref{EOM-AdS4K}) for $t(r)$ and its derivative $t'(r)$ are obtained
along the same lines as in $AdS_5$-Kasner. These are plotted in
Fig.~\ref{LL-i-AdS4K-Num}, which reveal that the behaviour of the
complexity surfaces $t(r)$ and their derivatives are similar to those
in $AdS_5$-Kasner.

\subsection{Holographic complexity of AdS$_7$-Kasner spacetime}
\label{C-AdS7K-subsec}

The equation of motion for the complexity surface $t(r)$ for
$AdS_7$-Kasner spacetime using $d_i=5$ in (\ref{EOM-t[r]-AdS-Kasner})
becomes
\begin{eqnarray}\label{EOM-AdS7K}
r\,t(r)\,t''(r)
-6t(r)\,t'(r) \left(1-t'(r)^2\right) + r \left(1-t'(r)^2\right) = 0\,.
\end{eqnarray}
As in $AdS_5$-Kasner spacetime, we solve this perturbatively and
numerically.\\

{\bf Perturbative results}: The perturbative solution of
(\ref{EOM-AdS7K}) for the ansatz $t(r)=t_0+ {\bf \sum_{n \in \mathbb{Z}_+}} c_n r^n$ is given as:
\begin{eqnarray}
\label{soln-t(r)-AdS7K}
& & \hskip -0.3in t(r)=t_0+\frac{r^2}{10 t_0}-\frac{23 r^4}{3000 t_0^3}. 
\end{eqnarray}
The solution (\ref{soln-t(r)-AdS7K}) for $t(r)$ and its derivative $t'(r)$
for $AdS_7$-Kasner are qualitatively similar to those in $AdS_5$-Kasner.

\begin{figure}[h]
\begin{subfigure}
  \centering
  \includegraphics[width=.5\linewidth]{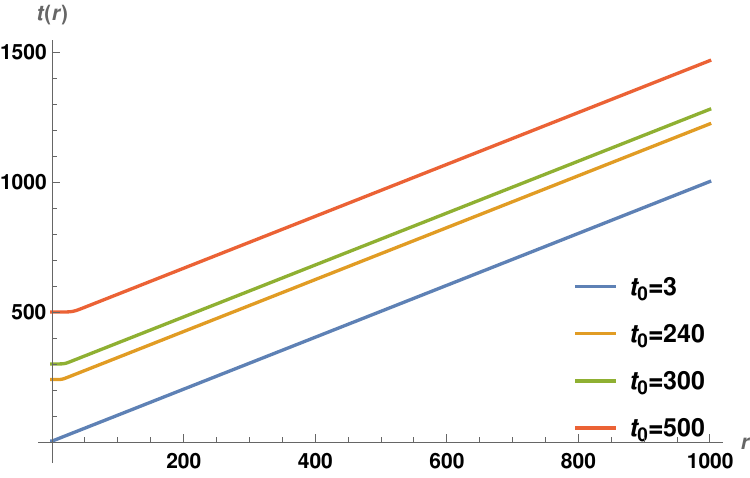}
\end{subfigure}
\begin{subfigure}
  \centering
  \includegraphics[width=.5\linewidth]{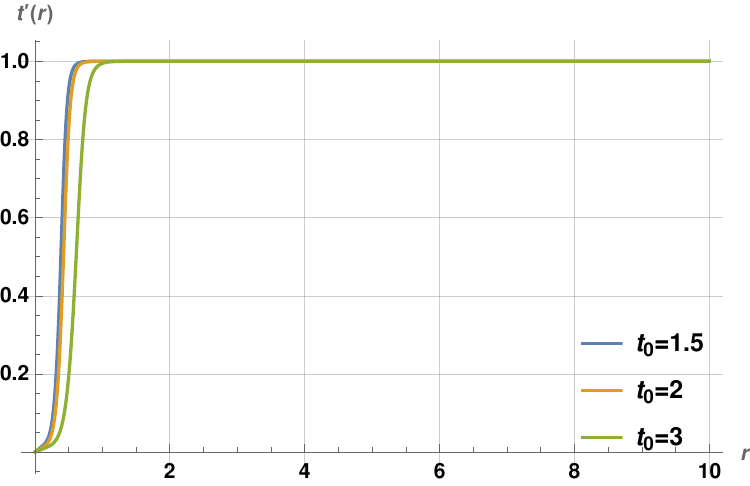}
\end{subfigure}
\begin{subfigure}
  \centering
  \includegraphics[width=.5\linewidth]{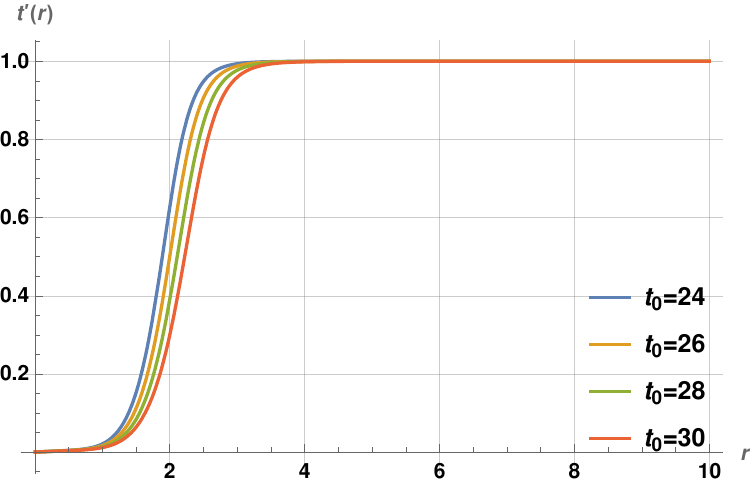}
\end{subfigure}
\begin{subfigure}
  \centering
  \includegraphics[width=.5\linewidth]{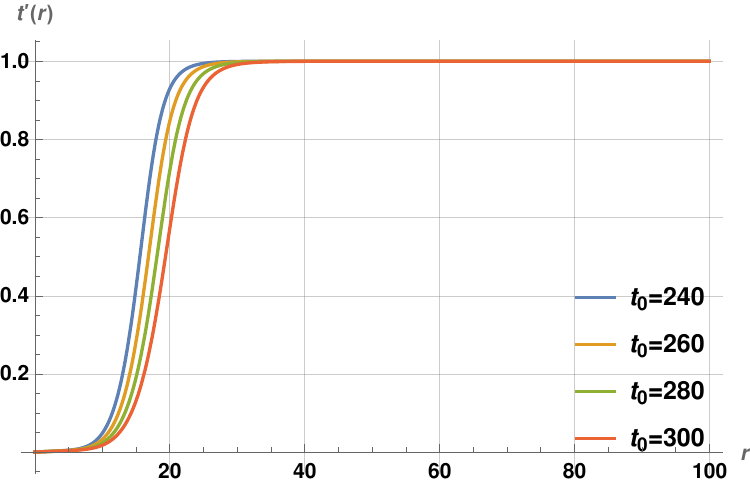}
\end{subfigure}
\caption{Plots of $t(r)$ with $r$ and $t'(r)$ with $r$ in $AdS_7$-Kasner
  spacetime for different $t_0$ slices.}
\label{LL-i-AdS7K-Num}
\end{figure}

{\bf Numerical results}: Using the above, we can pin down boundary
conditions near the boundary and then solve (\ref{EOM-AdS7K}) numerically.
This is similar to the analysis in $AdS_5$-Kasner and the solution $t(r)$
of (\ref{EOM-AdS7K}) and its derivative $t'(r)$ are plotted in
Fig.~\ref{LL-i-AdS7K-Num}. We see that the $AdS_7$-Kasner spacetime
gives similar results.

\subsection{Numerical computation of complexity, $AdS$-Kasner}
\label{NCADSK}

\begin{figure}[h]
\begin{subfigure}
  \centering
  \includegraphics[width=.45\linewidth]{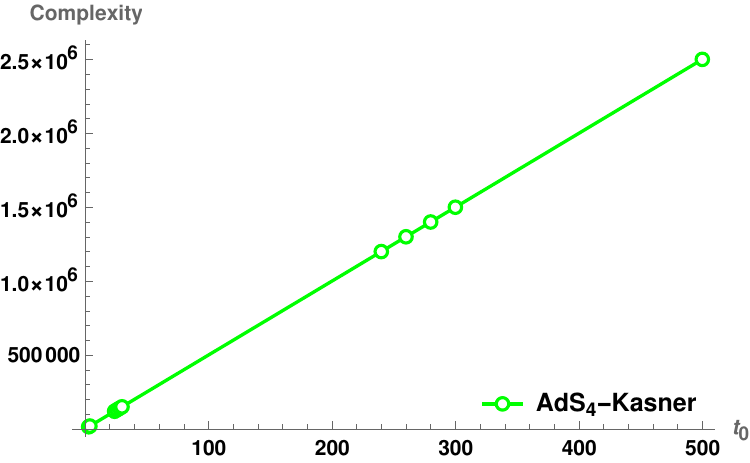}
\end{subfigure}
\begin{subfigure}
  \centering
  \includegraphics[width=.45\linewidth]{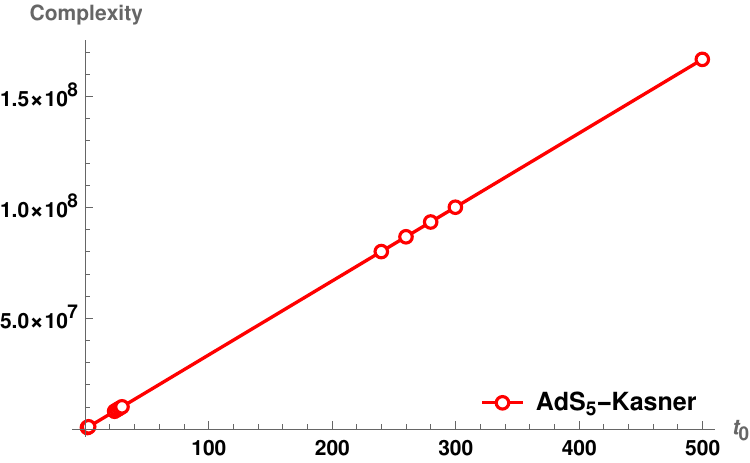}
\end{subfigure}
\begin{subfigure}
  \centering
  \includegraphics[width=.45\linewidth]{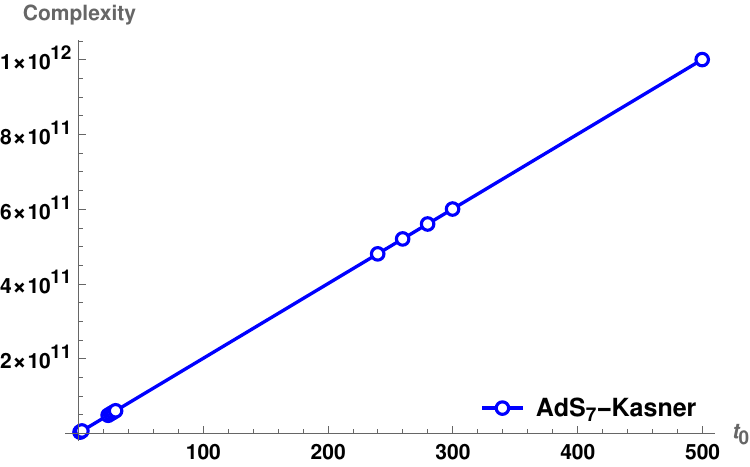}
\end{subfigure}
\caption{Numerical plots of holographic volume complexity with $t_0$ in
  $AdS_{4,5,7}$-Kasner spacetimes.}
\label{NC-AdSK}
\end{figure}

We evaluate holographic complexity of $AdS_{5,4,7}$-Kasner spacetime
numerically using the numerical solutions discussed above and performing
the numerical integration in $C$. The expression of holographic
complexity for $AdS_5$, $AdS_4$ and $AdS_7$-Kasner spacetimes are given as:
\begin{eqnarray}
\label{Complexity-behavior-NHM-AdS5-Kasner-NC}
({\rm AdS_{d_i+2}\ Kasner}) \ \ \ \ \ \ \ \
C = \frac{V_{d_i}{R^{d_i}}}{G_{d_i+2}} \int_\epsilon dr\,
\left( {t(r)\ \sqrt{1-t'(r)^2\,} \over r^{d_i+1}} \right)\,.\qquad
\end{eqnarray}
To perform the integrals numerically, we set the lengthscales
$V_{d_i},\ R^{d_i},\ G_{d_i+2}$ to unity and take $\epsilon=10^{-2}$ at
the lower end.
The upper end of the integration domain is irrelevant since the
complexity surfaces become lightlike eventually as $r$ increases so
the complexity integral has negligible contribution there, as stated
in sec.~\ref{sec:lightliket(r)}. Then as an order-of-magnitude
estimate, $t_0\sim 100$ gives
$C\sim {t_0\over \epsilon^{d_i}}\sim 10^{2d_i+2}$\ which corroborates
with the scales in Fig.~\ref{NC-AdSK} which displays the variation
of complexity with $t_0$ in $AdS$-Kasner spacetime. From the Figure,
we see that holographic volume complexity decreases linearly as the
anchoring time slice approaches the vicinity of the singularity,
\ie\ as $t_0\ra 0$. Thus the dual Kasner state appears to be of
vanishingly low complexity, independent of the reference state.



It is worth making a few comparisons on holographic complexity across
$AdS_{5,7,4}$ Kasner spacetimes, based on the numerical results in
Fig.~\ref{LL-i-t[r]},
Fig.~\ref{LL-i}\ ($AdS_5$-K),\ Fig.~\ref{LL-i-AdS7K-Num}\ ($AdS_7$-K),
and Fig.~\ref{LL-i-AdS4K-Num}\ ($AdS_4$-K). Relative to
$AdS_5$-Kasner, we see that the complexity surfaces become lightlike
at smaller $r$-values in $AdS_7$-Kasner.  Thus complexity of
$AdS_7$-Kasner acquires vanishing contributions at smaller $r$-values
relative to $AdS_5$-Kasner. Likewise, we see that the $AdS_5$-Kasner
displays the lightlike regime at smaller $r$-values than
$AdS_4$-Kasner. Thus complexity surfaces in higher dimensional
$AdS$-Kasner acquire lightlike regimes earlier than those in lower
dimensional ones. However it is clear that the leading divergence
behaviour is larger for higher dimensions, since the extremal codim-1
surface volumes have dominant contributions from the near boundary
region. With cutoff $\epsilon\equiv\Lambda_{UV}^{-1}$ we have the
scaling
\be\label{C-t0}
C\ \sim\ {R^{d_i+1}\over G_{d_i+2}\,R}\, {V_{d_i}\over\epsilon^{d_i}}\,t_0\
\equiv\ N_{dof}\, V_{d_i}\Lambda_{_{UV}}^{d_i}\, t_0\,,
\ee
reflecting the fact that complexity scales with the number of degrees
of freedom in the dual field theory and with spatial volume in units
of the UV cutoff. Some intuition for this can be obtained from the
form of the complexity volume functional
(\ref{Complexity-behavior-NHM-AdS5-Kasner-NC}) where the
$\sqrt{1-(t')^2}$ factor is amplified by the ${1\over
  r^{d_i}}$-factor. Both the spacelike part ($t'\ll 1$) of the
complexity surface and the transition to the lightlike part (where
$t'$ is changing) are amplified by the ${1\over r^{d_i}}$-factor to a
greater degree at larger $d_i$. Thus higher dimensional $AdS$-Kasner
hits the lightlike regime and vanishing complexity at smaller
$r$-values relative to lower dimensions.

Overall from (\ref{C-t0}) we see that
\be
{dC\over dt_0}\ \sim\ N_{dof}\, V_{d_i}\Lambda_{_{UV}}^{d_i}\,,
\ee
which arises from just the near-boundary UV part of the complexity
surface, with the lightlike part giving vanishing contributions.
This is consistent with complexity scaling as the number of microscopic
degrees of freedom in (a lattice approximation of) the CFT. In this
light, it appears that in the vicinity of the singularity, there is a
thinning of the effective number of degrees of freedom. As $t_0\ra 0$,
space entirely Crunches and there are no effective qubits, and so
correspondingly complexity vanishes.


\section{Complexity:\ hyperscaling violating cosmologies}\label{HSV-sec}

Various cosmological deformations of conformally $AdS$ or hyperscaling
violating theories were found in \cite{Bhattacharya:2020qil}\ (see
App.~\ref{sec:HolCosD}).
The 2-dim form of these backgrounds can be described in terms of the
2-dim dilaton gravity action (\ref{2ddg-action}) with dilaton
potential, parameters, as well as the $(t,r)$-scaling exponents in
(\ref{2d-tr-exp}) given below. Also appearing below is the higher
dimensional cosmology:
\bea\label{HDBC-ntheta}
&& 
U = 2\Lambda \phi^{1/d_i}\, e^{\gamma\Psi}\,,\qquad 2\Lambda=-(d_i+1-\theta)(d_i-\theta)\,,\qquad \gamma=\frac{-2 \theta}{\sqrt{2 d_i (d_i-\theta)(-\theta)}}\,,\nn\\
&& \quad ds^2=\frac{{R^{2}}r^{\frac{2 \theta}{d_i}}}{r^2}\Biggl(\frac{-dt^2+dr^2}{t^{\gamma \alpha}}+t^{2/d_i}dx_i^2 \Biggr), \quad e^{\Psi}=t^\alpha r^{\sqrt{2(d_i-\theta)\frac{(-\theta)}{d_i}}}\,,
\nn\\
&& k=1\,,\quad m=-(d_i-\theta)\,,\quad a={\al^2\over 2}\,,\quad 
\alpha=-\gamma \pm \sqrt{\gamma^2+\frac{2(d_i-1)}{d_i}}\,,\nn\\
&& b=-{(d_i-\theta)(d_i+1)\over d_i}\,,\quad
\beta=\sqrt{{2(d_i-\theta)(-\theta)\over d_i}}\,.
\eea
With $\al$ taken positive, $e^\Psi\ra 0$ as $t\ra 0$ and we obtain
\be\label{hv-al>0}
\al=-\gamma + \sqrt{\gamma^2+\frac{2(d_i-1)}{d_i}} \quad\ra\quad
a=\left(\sqrt{\frac{d_i-\theta -1}{d_i-\theta }}
-\sqrt{\frac{(-\theta) }{d_i (d_i-\theta)}}\right)^2 .
\ee
In this section, we restrict to  Lorentz invariance: the Lifshitz
exponent is $z=1$. Then the null energy conditions \cite{Dong:2012se}
implies that the hyperscaling violating exponent $\theta$ is constrained
as
\be
(d_i-\theta)(d_i(z-1)-\theta) \geq 0\,,\qquad
(z-1)(d_i+z-\theta) \geq 0 \quad\xrightarrow{z=1}\quad \theta\leq 0\,.
\ee
The other possibility $\theta>d_i$ has undesirable properties suggesting
instabilities \cite{Dong:2012se}.

Time-independent backgrounds of this sort appear in the dimensional
reduction \cite{Dong:2012se} over the transverse spheres of
nonconformal $Dp$-branes \cite{Itzhaki:1998dd}, and the
$\theta$-exponent is then related to the nontrivial running of the
gauge coupling. Reductions of nonconformal $Dp$-branes over the
transverse spheres and over the brane spatial dimensions leads to
2-dim dilaton gravity theories \cite{Kolekar:2018chf} with dilaton
potentials as in (\ref{HDBC-ntheta}) above, and the 2-dim
dilaton then leads to a holographic $c$-function encoding the
nontrivial renormalization group flows. Some of the analysis there, as
well as in \cite{Dong:2012se}, may be helpful to keep in mind in our
discussions here. In particular, the D2-brane and D4-brane supergravity
phases give rise to\ $d_i=2,\ \theta=-{1\over 3}$\ and\
$d_i=4,\ \theta=-1$,\ respectively, both with $z=1$. In these cases,
the Big-Bang/Crunch singularities may be interpreted as appropriate
Kasner-like deformations of the nonconformal $Dp$-brane backgrounds,
although again the time-dependence does not switch off asymptotically
with corresponding difficulties in interpretation as severe
time-dependent deformations of some vacuum state.

We will focus on these in studying the Big-Bang/Crunch hyperscaling
violating cosmological backgrounds in (\ref{HDBC-ntheta}) above and
study holographic complexity thereof. The calculations are broadly similar
to those in $AdS$ Kasner spacetimes earlier, but with
interesting detailed differences.

Using the exponents in (\ref{HDBC-ntheta}), (\ref{hv-al>0}), the
holographic volume complexity (\ref{Complexity-behavior-NHM}) simplifies
to
\be\label{Complexity-behavior-NHM-hyperscaling-violating}
C =\frac{V_{d_i}{R^{d_i}}}{G_{d_i+2}} \int_\epsilon dr \left(r^{\frac{(d_i+1) (\theta -d_i)}{d_i}}  t(r)^{\frac{1}{2}
   \left(\left(\sqrt{\frac{d_i-\theta -1}{d_i-\theta }}-\sqrt{\frac{(-\theta) }{d_i (d_i-\theta)}}\right)^2+\frac{1}{d_i}+1\right)} \ \sqrt{1-t'(r)^2} \right).
\ee
Extremizing with $t(r)$, we obtain the Euler-Lagrange
equation of motion for the complexity surface $t(r)$ as
\begin{eqnarray}
\label{t[r]-EOM-NCB}
& & 
r \left(d_i \left(\sqrt{\frac{d_i-\theta -1}{d_i-\theta }}-\sqrt{\frac{(-\theta )}{d_i
   (d_i-\theta )}}\right)^2+d_i+1\right) \left(1-t'(r)^2\right) \nonumber\\
& & \qquad\quad  +\ 2 d_i\,r\,t(r)\,t''(r)
-2 (d_i+1) (d_i-\theta)\,t(r)\,t'(r) \left( 1-t'(r)^2\right)
= 0\,.
\end{eqnarray}
In the semiclassical limit ($t'(r)\ll 1$), we can ignore terms like $t'(r)^2, t'(r)^3$ in the equation of motion (\ref{t[r]-EOM-NCB}), which then 
simplifies to
\begin{eqnarray}
\label{t[r]-EOM-NCB-i}
& & \hskip -0.4in t(r) \left(2 (d_i+1) (d_i-\theta ) t'(r)-2 d_i r t''(r)\right)-r \left(d_i \left(\sqrt{\frac{d_i-\theta -1}{d_i-\theta
   }}-\sqrt{\frac{(-\theta )}{d_i (d_i-\theta )}}\right)^2+d_i+1\right)=0.\nonumber\\
\end{eqnarray}
Now, we solve equations (\ref{t[r]-EOM-NCB}) and (\ref{t[r]-EOM-NCB-i})
perturbatively using ansatze similar to those in the $AdS$-Kasner
spacetime, \ie\ $t(r)=t_0+ \sum_{n \in \mathbb{Z}_+} c_n r^n$\,. We
illustrate this in detail by analysing holographic volume complexity
for two cases:
(i) $d_i=2,\ \theta=-1/3$ in sec.~\ref{D2}, and (ii) $d_i=4,\ \theta=-1$
in sec.~\ref{D4}. Analysing other cases reveals similar results. To
differentiate between the different solutions, we will use different
coefficients for the different cases, e.g., $g_n$, $s_n$ etc.



\subsection{$d_i=2, \ \theta=-{1\over 3}$}\label{D2}

This case is related to the $D2$-brane supergravity phase as stated
earlier, and we analyze the perturbative and numerical solutions now.
 
The equation of motion (\ref{t[r]-EOM-NCB}) with exponents
(\ref{HDBC-ntheta}), (\ref{hv-al>0}), for this case simplifies to
\begin{eqnarray}
\label{EOM-Num-d=2-theta=-1by3}
14 r\,t(r)\,t''(r)  
- \left(2 \sqrt{2}-15\right) r \left(1-t'(r)^2\right)
-49 t(r)\,t'(r) \left( 1-t'(r)^2\right)=0\,.
\end{eqnarray}
The solution $t(r)$ up to ${\cal O}(r^{4})$ is given as
\begin{eqnarray}
\label{t[r]-general-soln-NCB-IR-theta=-1by3}
t(r) = t_0+g_2 r^2+g_4 r^4\,,
\end{eqnarray}
with $g_i$ in (\ref{ci's-theta=-1by3-general}). The behaviour of the
complexity surfaces (\ref{t[r]-general-soln-NCB-IR-theta=-1by3}) with
$r$ for different $t_0$ values is qualitatively similar to those in
$AdS$-Kasner (see Fig.~\ref{t[r]-plots}) so we will not display the plots.

When $t'(r)\ll 1$, we can ignore the higher order terms in 
(\ref{EOM-Num-d=2-theta=-1by3}) to obtain
\begin{eqnarray}
\label{EOM-Num-d=2-theta=-1by3-i}
2 \left(2 \sqrt{2}-15\right) r -14 t(r) \left(2 r t''(r)-7 t'(r)\right)=0.
\end{eqnarray}
This has solution up to ${\cal O}(r^{4})$ given as
\begin{eqnarray}
\label{t[r]-soln-NCB-IR-theta=-1by3}
t(r) = t_0+s_2 r^2+s_4 r^4\,,
\end{eqnarray}
with $s_i$ in (\ref{ci's-theta=-1by3}). The behaviour of
(\ref{t[r]-soln-NCB-IR-theta=-1by3}) is qualitatively similar to
that in $AdS$-Kasner.\par

\begin{figure}[h]
\begin{subfigure}
  \centering
  \includegraphics[width=.5\linewidth]{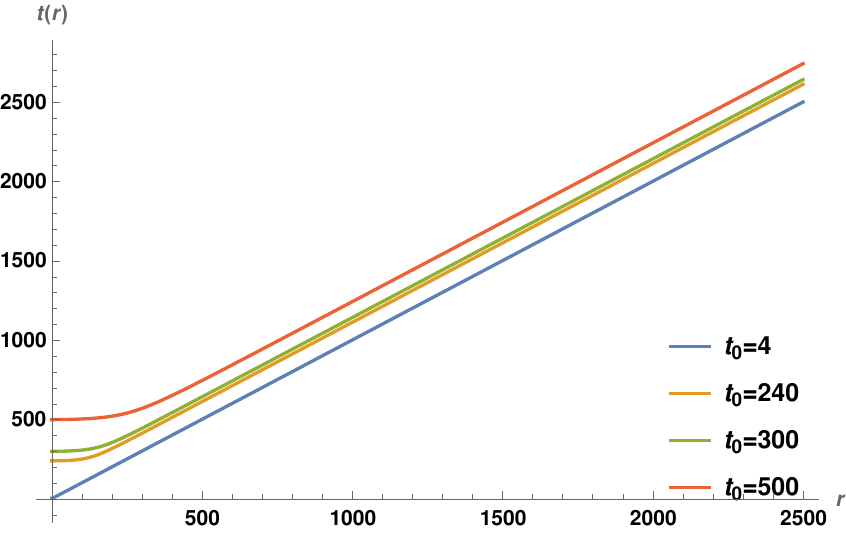}
\end{subfigure}
\begin{subfigure}
  \centering
  \includegraphics[width=.5\linewidth]{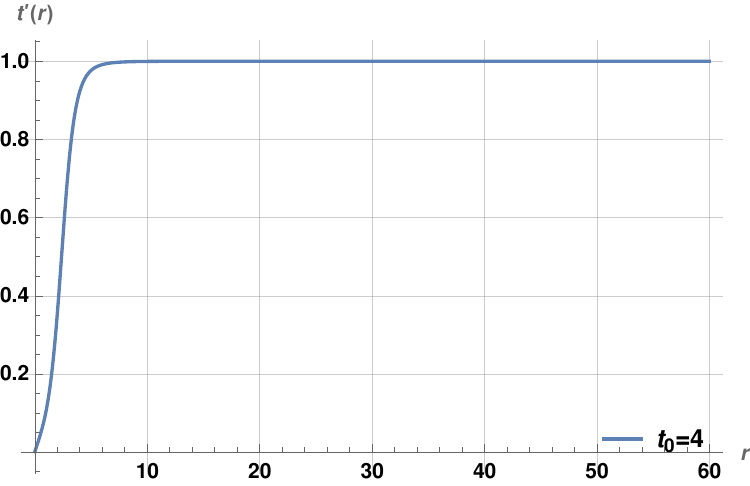}
\end{subfigure}
\begin{subfigure}
  \centering
  \includegraphics[width=.5\linewidth]{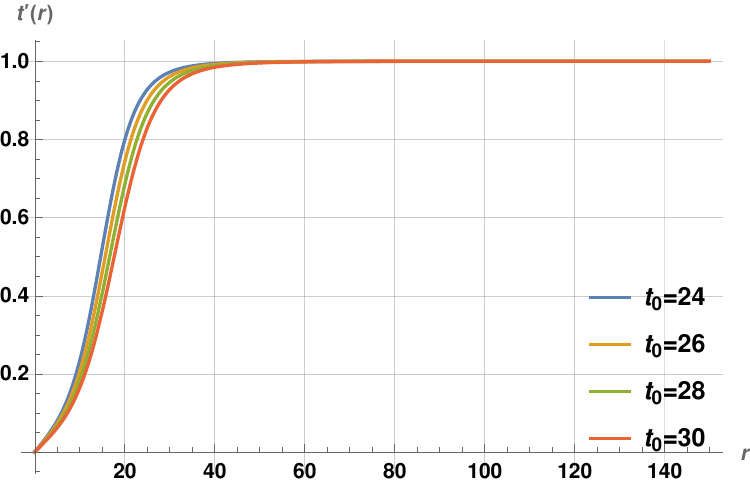}
\end{subfigure}
\begin{subfigure}
  \centering
  \includegraphics[width=.5\linewidth]{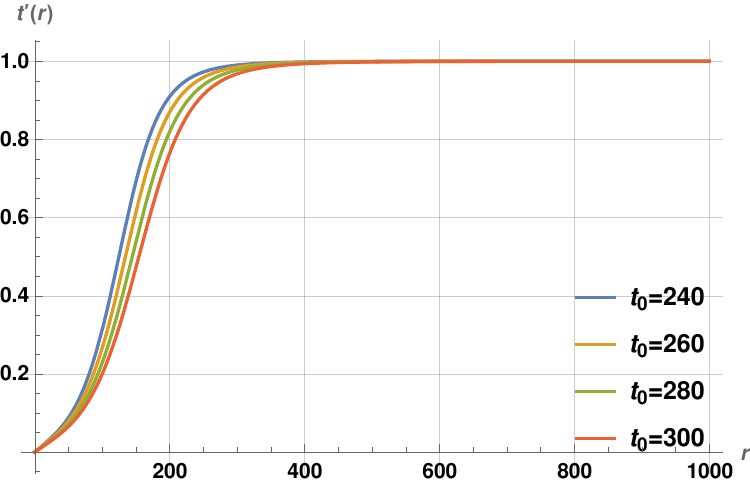}
\end{subfigure}
\caption{Numerical plots of $t(r)$ versus $r$ and $t'(r)$ versus $r$ for $d_i=2$ and $\theta=-1/3$ in hv cosmology.}
\label{LL-i-HSV}
\end{figure}

Solving (\ref{EOM-Num-d=2-theta=-1by3}) numerically along similar lines
as in $AdS$-Kasner, we obtain the variation of the complexity surfaces
and their derivatives with $r$: this is shown in Fig.~\ref{LL-i-HSV}.


{\bf Holographic complexity}: With $d_i=2,\ \theta=-{1\over 3}$ in
(\ref{Complexity-behavior-NHM-hyperscaling-violating}) gives
\begin{eqnarray}
\label{Complexity-behavior-NHM-hyperscaling-violating-di=2-theta=-1by3}
C =\frac{V_2 {R^{2}}}{G_{4}} \int_\epsilon dr \left(\frac{t(r)^{\frac{15}{14}-\frac{\sqrt{2}}{7}} \sqrt{1-t'(r)^2}}{r^{7/2}} \right).
\end{eqnarray}
Restricting to the regime $t'(r)\ll 1$, we perform this integral 
rewriting as
\begin{eqnarray}
\label{Complexity-behavior-NHM-hyperscaling-violating-di=2-theta=-1by3-i}
C =\frac{V_2 {R^{2}}}{G_{4}} \int_\epsilon^{r_\Lambda} dr \left(\frac{t(r)^{\frac{15}{14}-\frac{\sqrt{2}}{7}} \left({1-\frac{t'(r)^2}{2}}\right)}{r^{7/2}} \right) .
\end{eqnarray}
Then using
(\ref{t[r]-soln-NCB-IR-theta=-1by3}) truncating to\
$t(r)=t_0+s_2 r^2+s_4 r^4$ using (\ref{ci's-theta=-1by3}), the complexity
(\ref{Complexity-behavior-NHM-hyperscaling-violating-di=2-theta=-1by3-i})
up to next-to-leading-order in $t_0$ is obtained as
\begin{eqnarray}
\label{Complexity-behavior-NHM-hyperscaling-violating-di=2-theta=-1by3-iii}
C =\frac{V_2 {R^{2}}}{G_{4}} \left(\frac{2}{5} t_0^{\frac{15}{14}-\frac{\sqrt{2}}{7}} \left(\frac{1}{\epsilon
   ^{5/2}}-\frac{1}{r_\Lambda^{5/2}}\right) -\frac{2883 \left(\frac{1}{t_0}\right)^{\frac{13}{14}+\frac{\sqrt{2}}{7}} 
   \left(\frac{1}{\sqrt{r_\Lambda}}-\frac{1}{\sqrt{\epsilon }}\right)}{50 \left(233+60 \sqrt{2}\right)}\right).
\end{eqnarray}


\subsection{$d_i=4, \ \theta=-1$} \label{D4}

This is related to the $D4$-brane supergravity phase as stated earlier.

\begin{figure}[h]
\begin{subfigure}
  \centering
  \includegraphics[width=.5\linewidth]{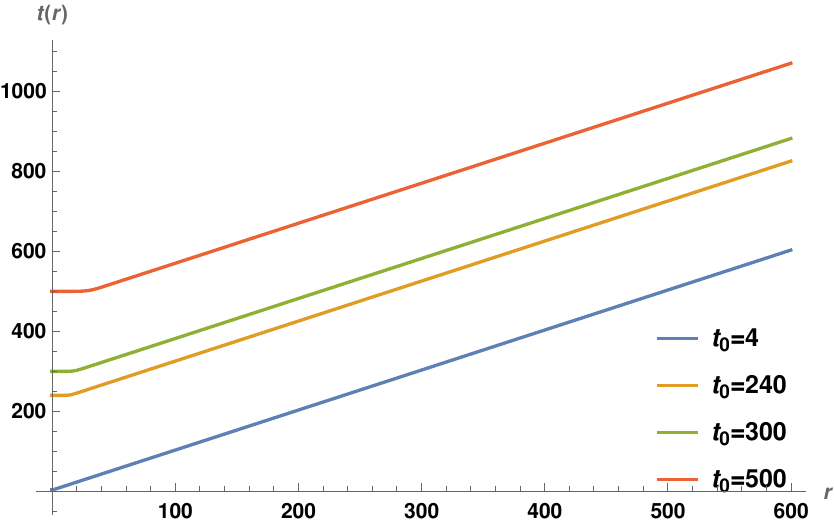}
\end{subfigure}
\begin{subfigure}
  \centering
  \includegraphics[width=.5\linewidth]{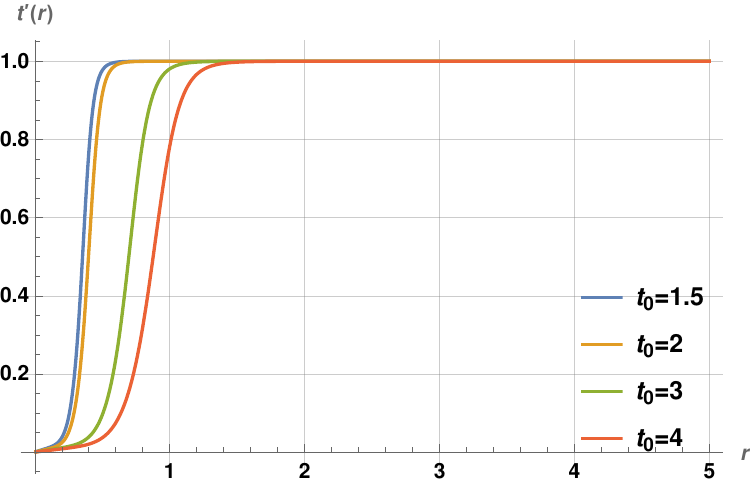}
\end{subfigure}
\begin{subfigure}
  \centering
  \includegraphics[width=.5\linewidth]{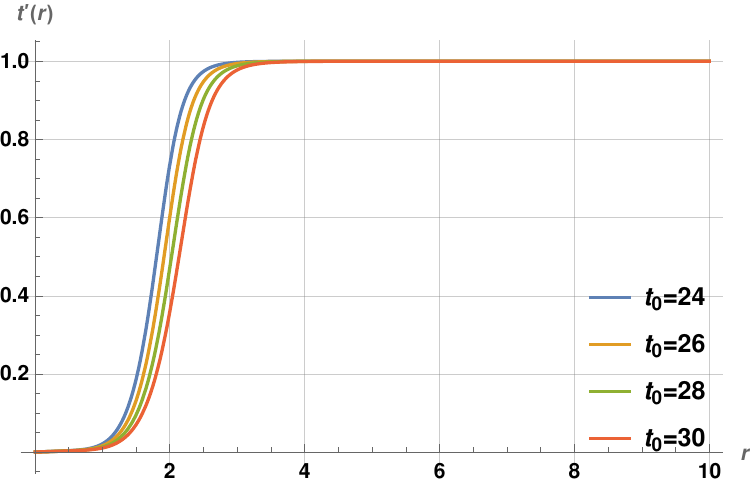}
\end{subfigure}
\begin{subfigure}
  \centering
  \includegraphics[width=.5\linewidth]{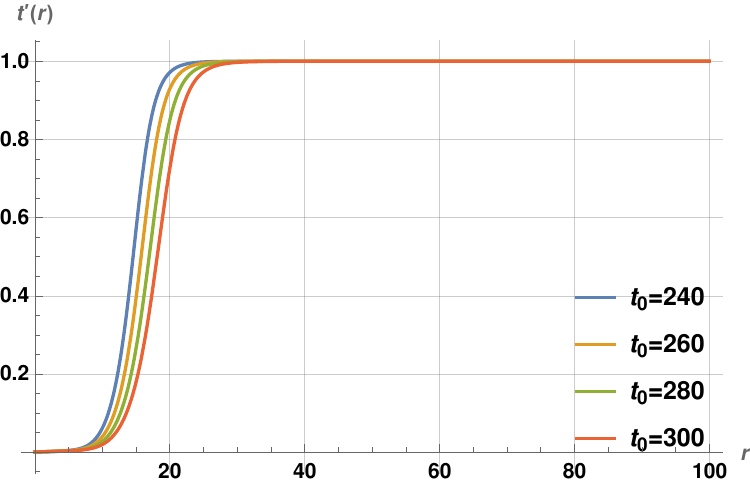}
\end{subfigure}
\caption{Plots of complexity surface versus $r$ and $t'(r)$ versus $r$ for $d_i=4$ and $\theta=-1$ in hyperscaling violating cosmology.}
\label{LL-i-HSV-theta=-1}
\end{figure}

For $d_i=4$ and $\theta=-1$, the equation of motion (\ref{t[r]-EOM-NCB})
becomes
\begin{eqnarray}
\label{EOM-Num-d=4-theta=-1}
20\, r\,t(r)\,t''(r) + 17\, r \left(1-t'(r)^2\right) 
-125\, t(r)\,t'(r) \left(1-t'(r)^2\right)=0\,.
\end{eqnarray}
Using the ansatz $t(r)=t_0 + \sum_{n \in \mathbb{Z}_+} y_n r^n$, we obtain
the following solution to (\ref{EOM-Num-d=4-theta=-1}):
\begin{eqnarray}
\label{t[r]-general-soln-NCB-IR-theta=-1}
t(r) = t_0+y_2 r^2+y_4 r^4\,,
\end{eqnarray}
with $y_i$ given in (\ref{ci's-theta=-1-general}). The plots of the
solution (\ref{t[r]-general-soln-NCB-IR-theta=-1}) and its derivatives
are qualitatively similar to those in $AdS$-Kasner spacetimes and the
$d_i=2, \theta=-{1\over 3}$ hv-cosmology.\ Ignoring higher order terms
in (\ref{EOM-Num-d=4-theta=-1}), we obtain
\begin{eqnarray}
\label{EOM-Num-d=4-theta=-1-ii}
-34 r +5 t(r) \left(-8 r t''(r)+50 t'(r)\right)=0\,.
\end{eqnarray}
Solving this with the ansatz $t(r)=t_0 + \sum_{n \in \mathbb{Z}_+} v_n r^n$
gives the qualitatively similar complexity surface (with
$v_i$ in (\ref{ci's-theta=-1})):
\begin{eqnarray}
\label{t[r]-soln-NCB-IR-theta=-1}
t(r) = t_0+v_2 r^2+v_4 r^4\,.
\end{eqnarray}
Solving the nonlinear equation (\ref{EOM-Num-d=4-theta=-1}) numerically
as in $AdS$-Kasner gives numerical solutions for the complexity surfaces
$t(r)$ and their derivatives. These are shown in Fig.~\ref{LL-i-HSV-theta=-1}.

{\bf Holographic complexity}:\ Substituting $d_i=4, \ \theta=-1$ in
(\ref{Complexity-behavior-NHM-hyperscaling-violating}) gives
\begin{eqnarray}
\label{Complexity-behavior-NHM-hyperscaling-violating-di=4-theta=-1}
C =\frac{V_4 {R^{4}}}{G_6 } \int_\epsilon dr \left(\frac{t(r)^{17/20} \sqrt{1-t'(r)^2}}{r^{25/4}}\right)\ \sim\ \frac{V_4 {R^{4}}}{G_6} \int_\epsilon^{r_{\Lambda}} dr \left(\frac{t(r)^{17/20} \left(1-\frac{t'(r)^2}{2}\right)}{r^{25/4}}\right).
\end{eqnarray}
Similar to the discussion for $d_i=2, \theta=-{1\over 3}$, we approximate
as $t(r)=t_0+v_2 r^2+v_4 r^4$\ (with $v_2$, $v_4$ in (\ref{ci's-theta=-1}))
and simplify (\ref{Complexity-behavior-NHM-hyperscaling-violating-di=4-theta=-1}).
The calculations are similar to the earlier case: complexity up to
next-to-leading-order in $t_0$ is
\begin{eqnarray}
\label{Complexity-behavior-NHM-hyperscaling-violating-di=4-theta=-1-i}
& & \hskip -0.5in
C =\frac{V_4 {R^{4}}}{G_6} \Biggl[t_0^{17/20} \left(\frac{4}{21 \epsilon ^{21/4}}-\frac{4}{21
   {r_\Lambda}^{21/4}}\right) \nonumber\\
   & &\hskip -0.3in +\left(\frac{1}{t_0}\right)^{23/20} \left(\frac{4913}{525
   \left(\sqrt{715}-13\right) \left(13+\sqrt{715}\right) \epsilon ^{13/4}}-\frac{4913}{525
   \left(\sqrt{715}-13\right) \left(13+\sqrt{715}\right) {r_\Lambda}^{13/4}}\right)\Biggr].\nonumber\\
   \end{eqnarray}

{\bf \{$d_i=2, \ \theta=-{1\over 3}$\}\ vs\ \{$d_i=4, \ \theta=-1$\}\
  hv-cosmologies}:\ \
Comparing Fig.~\ref{LL-i-HSV} and Fig.~\ref{LL-i-HSV-theta=-1}, 
we see that the numerical solution of $d_i=4, \ \theta=-1$ hv-cosmology becomes lightlike for a smaller $r$-value relative to that for $d_i=2, \ \theta=-{1\over 3}$\,. The $\sqrt{1-t'(r)^2}$ factor in the volume complexity expression
implies $d_i=4, \ \theta=-1$ complexity vanishing earlier relative to
that in $d_i=2, \ \theta=-{1\over 3}$\,. These theories have effective
space dimension $d_{eff}=d_i-\theta$ so the effective dimensions are
$d_{eff}=5$ and $d_{eff}={7\over 3}$ respectively: so the larger
effective dimension case acquires vanishing complexity at smaller
$r$-values. This is similar to the observations in $AdS$-Kasner
discussed in sec.~\ref{NCADSK}.

Let us summarize the results for complexity in $AdS_5$-Kasner and
hyperscaling violating cosmologies up to next-to-leading-order in $t_0$\
(the results to this order are the same whether we use the linearized
equation ignoring higher order terms in $t'(r)$, or the full nonlinear
one):
\begin{eqnarray}
\label{summary-AdS5K-hvLif-cosmo}
& & \hskip -0.3in ({\rm AdS_5-Kasner}) \ \ \ \ \ \ \ \ C \approx \frac{V_3 {R^{3}}}{G_{5}}\Biggl[t_0 \left(\frac{0.3}{\epsilon ^3}-\frac{0.3}{r_\Lambda^3}\right)+\frac{1}{t_0}\left(\frac{0.1}{\epsilon
   }-\frac{0.1}{r_\Lambda}\right)+O\left(\left(\frac{1}{t_0}\right)^3\right)\Biggr],\nonumber\\
& & \hskip -0.3in \left(d_i=2, \ \theta=-{1\over 3}\right) \ \ \ \  C \approx \frac{V_2 {R^{2}}}{G_{4}} \left(\frac{2}{5} t_0^{\frac{15}{14}-\frac{\sqrt{2}}{7}} \left(\frac{1}{\epsilon
   ^{5/2}}-\frac{1}{r_\Lambda^{5/2}}\right) -\frac{2883 \left(\frac{1}{t_0}\right)^{\frac{13}{14}+\frac{\sqrt{2}}{7}} 
   \left(\frac{1}{\sqrt{r_\Lambda}}-\frac{1}{\sqrt{\epsilon }}\right)}{50 \left(233+60 \sqrt{2}\right)}\right),\nn\\
& &  \hskip -0.3in (d_i=4, \ \theta=-1) \ \ \ \ \ \ \
C \approx \frac{V_4 {R^{4}}}{G_6} \Biggl[t_0^{17/20} \left(\frac{4}{21 \epsilon ^{21/4}}-\frac{4}{21
   {r_\Lambda}^{21/4}}\right) \nonumber\\
   & &\hskip -0.3in +\left(\frac{1}{t_0}\right)^{23/20} \left(\frac{4913}{525
   \left(\sqrt{715}-13\right) \left(13+\sqrt{715}\right) \epsilon ^{13/4}}-\frac{4913}{525
   \left(\sqrt{715}-13\right) \left(13+\sqrt{715}\right) {r_\Lambda}^{13/4}}\right)\Biggr].\nonumber\\
\end{eqnarray}
The leading divergence of complexity from above is\
$C =\frac{V_2 {R^{2}}}{G_{4} \epsilon^{5/2}}\, t_0^{\frac{15-2\sqrt{2}}{14}}$\
for $d_i=2, \theta=-{1\over 3}$,\ and\
$C =\frac{V_4 {R^{4}}}{G_6 \epsilon^{21/4}}\, t_0^{17/20}$\ for 
$d_i=4, \theta=-1$.
Overall, at leading order in (\ref{summary-AdS5K-hvLif-cosmo}), we see
that the holographic complexity of AdS$_5$-Kasner spacetime is linearly
proportional to $t_0$ whereas in hyperscaling violating cosmologies,
complexity is proportional to  $t_0^{0.9}$ and $t_0^{0.85}$ summarized below:
\begin{eqnarray}
\label{C-Summarized-LO}
&& \qquad\qquad\qquad\qquad
({\rm AdS_5-Kasner}) \ \ \ \ \ \ \ \ \ \ \ C \propto t_0\,,  \nonumber\\
&& \left(d_i=2, \ \theta=-{1\over 3}\right) \ \ \ \ \ \ \ \ C \propto t_0^{0.9}\,,\qquad
\qquad  (d_i=4, \ \theta=-1) \ \ \ \ \ \ \  \ \ \ \  C \propto t_0^{0.85}\,.
\qquad
\end{eqnarray}
The complexity scaling in hv-cosmologies reflects the fact that the
dual theories live in some sort of effective space dimension 
$d_{eff}=d_i-\theta$. It might be interesting to understand the
underlying effective lattice qubit models simulating this behaviour
(these are perhaps distinct from relativistic CFTs, in light of the
general arguments in \cite{Susskind:2014rva}).

\subsection{Numerical computation of complexity in hv cosmologies}
\label{NChvLif}
\begin{figure}[h]
\begin{subfigure}
  \centering
  \includegraphics[width=.5\linewidth]{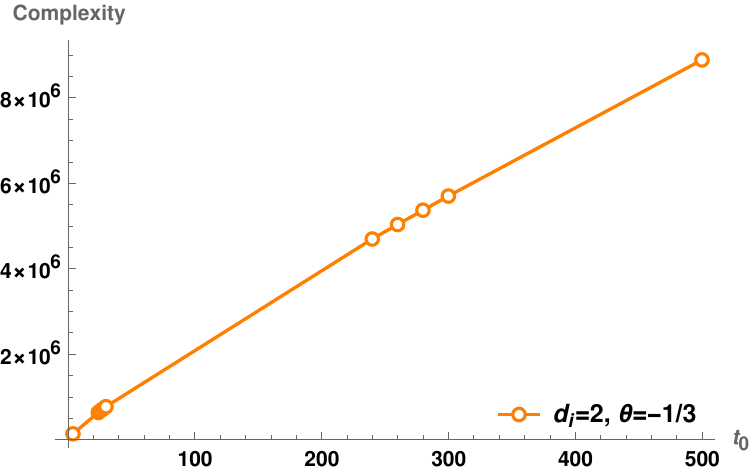}
\end{subfigure}
\begin{subfigure}
  \centering
  \includegraphics[width=.5\linewidth]{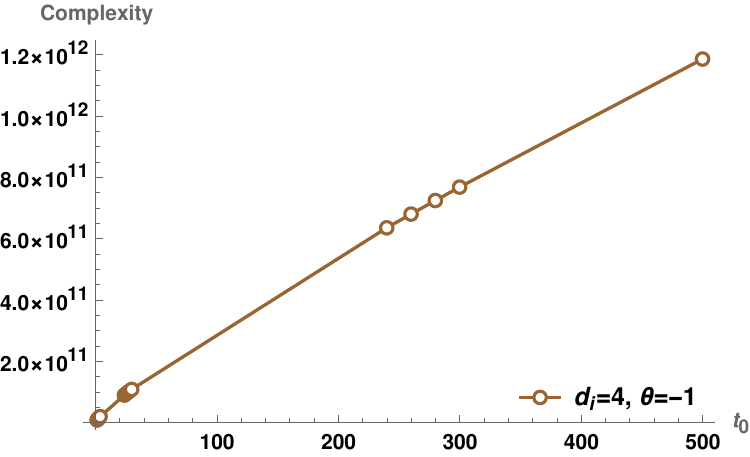}
\end{subfigure}
\caption{Numerical plots of holographic complexity versus $t_0$ for $d_i=2, \ \theta=-1/3$ and $d_i=4, \ \theta=-1$ cosmologies.}
\label{NC-hvLif-g}
\end{figure}
We now compute holographic complexity of hyperscaling violating
cosmologies, for $d_i=2, \ \theta=-{1\over 3}$, and $d_i=4, \ \theta=-1$,
as for $AdS$-Kasner spacetimes in sec.~\ref{NCADSK}. For this purpose,
we use the numerical solutions of the cosmologies as
discussed earlier, and numerically perform the integrations appearing
in the complexity expressions in
(\ref{Complexity-behavior-NHM-hyperscaling-violating-di=2-theta=-1by3})
for $d_i=2, \ \theta=-{1\over 3}$,\ and\
(\ref{Complexity-behavior-NHM-hyperscaling-violating-di=4-theta=-1})
for $d_i=4, \ \theta=-1$ (using the full nonlinear expression). The
variation of holographic complexity with $t_0$ in these backgrounds
is shown in Fig.~\ref{NC-hvLif-g}.

As we see in (\ref{C-Summarized-LO}), the complexity of hyperscaling
violating cosmologies does not scale linearly with $t_0$, unlike 
that in $AdS$-Kasner spacetimes. However the exponents are positive
so complexity continues to decrease with $t_0$ moving towards the
singularity, becoming vanishingly small as $t_0\ra 0$. Thus the dual
Kasner state continues to exhibit low complexity, as displayed in
Fig.~\ref{NC-hvLif-g}.

\section{Complexity:\ isotropic Lifshitz Kasner cosmologies}\label{Lif-cosmo-sec}

The 2-dim dilaton gravity formulation \cite{Bhattacharya:2020qil} led to new 
Kasner cosmologies with Lifshitz asymptotics. The equations of motion
are rather constraining however admitting only certain values for the
various exponents: in particular the cosmological solutions turn out
to have\ $\theta=0$ so $\gamma=0$ with Lifshitz exponent
$z=d_i$. In 2-dim form, they are described by the 2-dim dilaton
gravity action (\ref{2ddg-action}) with dilaton potential, parameters,
and the $(t,r)$-scaling exponents in (\ref{2d-tr-exp}) below. Also
given below is the higher dimensional Lifshitz Kasner cosmology:
\bea\label{LifCosExp}
&& U=\phi^{1/d_i} \Big(-c_1+{c_2\over\phi^2} e^{\lambda\Psi}\Big)\,,\quad
\lambda=-\sqrt{{2d_i\over d_i-1}}\,,\quad
c_1={2(2d_i-1)\over d_i}\,\quad c_2={2(d_i-1)\over d_i}\,, \nn\\ [1mm]
&& \qquad\qquad ds^2 = R^2\left(-{dt^2\over r^{2z}}+{dr^2\over r^2} + t^{2/d_i} {dx_i^2\over r^2}\right) ,
\qquad  e^\Psi=t^\al r^{-\al}\,, \quad z=d_i\,,\\
&& k=1,\quad m=-1\,,\quad a={d_i-1\over d_i}\,,\quad
b=-3+{1\over d_i}\,,\quad 
-\beta=\al=-\sqrt{{2(d_i-1)\over d_i}}\,.\ \ \nn
\eea
Here $R$ is the analog of the $AdS$ scale, and we are suppressing
an additional scale in $g_{tt}$ arising due to the nontrivial Lifshitz
scaling.
The nonrelativistic time-space scaling implies that lightlike
trajectories have\ $dt^2=r^{2z-2}dr^2$ so to identify lightlike limits
it is convenient to use $(t,\rho)$ coordinates with $\rho\sim r^z$. To
illustrate this and study complexity, we will focus on the Lifshitz
Kasner cosmology with $z=d_i=2$ and exponents $a={1\over 2}$\,,\
$b=-{5\over 2}$\, with metric
\begin{equation}\label{lifshitz_metric_z=2,d=2, theta=0_t,r_coords}
ds^2= {R^2}\left(-\frac{dt^2}{r^4} + \frac{dr^2}{r^2}+\frac{t}{r^2}\left(dx_{1}^2+dx_{2}^2\right)\right).
\end{equation}
Redefining $r^2\sim\rho$ and appropriately absorbing numerical factors
redefining the various lengthscales makes lightlike trajectories have
$dt^2=d\rho^2$, with the metric
(\ref{lifshitz_metric_z=2,d=2, theta=0_t,r_coords}) recast as 
\begin{equation}\label{lifshitz_metric_z=2,d=2, theta=0_t,rho_coords}
  ds^2 = {R^2}\left(-\frac{dt^2}{\rho^2}+\frac{d\rho^2}{\rho^2}+
  \frac{t}{\rho}(dx_1^2+dx_2^2)\right).
\end{equation}
Parametrizing the complexity surface by $t(\rho)$ gives the complexity
volume functional 
\begin{equation} \label{lifshitz_metric_z=2,d=2, theta=0_volume functional}
C= \frac{V_2 {R^3}}{G_4 R} \int_\epsilon d\rho \left(\frac{t(\rho)}{\rho^2} \sqrt{1-(t'(\rho))^2}\right),
\end{equation}
with $V_2=\int dx_1dx_2$. Extremizing for the complexity surface $t(\rho)$
gives the equation of motion 
\begin{equation}\label{lifshitz_metric_z=2,d=2, theta=0_volume functional_eom_t_rho_coords}
\rho\,t(\rho)\,t''(\rho)
-2 t(\rho)\,t'(\rho)(\left(1-t'(\rho)^2\right)
+ \rho\left(1-t'(\rho)^2\right) = 0 \, .
\end{equation}
The perturbative solution of (\ref{lifshitz_metric_z=2,d=2, theta=0_volume functional_eom_t_rho_coords}) for the ansatz
$t(\rho)=\sum_{n \in \mathbb{Z}_+} c_n \rho^n$ similar to the previous cases
up to ${\cal O}(\rho^{4})$ is given by
{
  \begin{eqnarray} \label{lifshitz_metric_z=2,d=2, theta=0_pertubative_sol_t}
& & \hskip -0.2in t(\rho)={t_0}+\frac{\rho^2}{2 {t_0}}-\frac{\rho^4}{8
   {t_0}^3}\,. \ \ \  
  \end{eqnarray}
}
The behavior of this perturbative solution is qualitatively similar to
that in $AdS$-Kasner and hv-cosmologies, so we suppress these plots here.
\begin{figure}[h]
  \begin{subfigure}
    \centering
    \includegraphics[width=.5\linewidth]{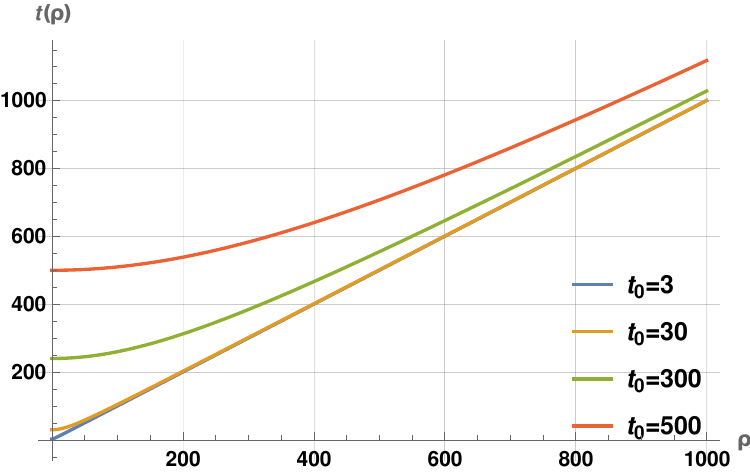}
  \end{subfigure}%
  \begin{subfigure}
    \centering
    \includegraphics[width=.5\linewidth]{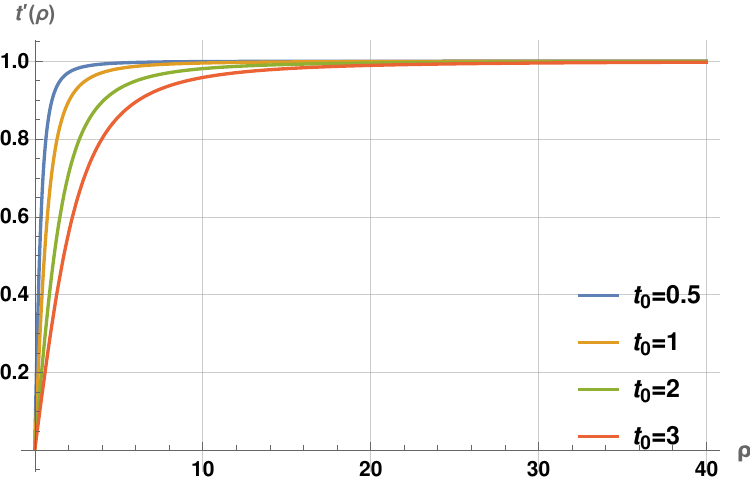}
  \end{subfigure}
 \begin{subfigure}
    \centering
    \includegraphics[width=.5\linewidth]{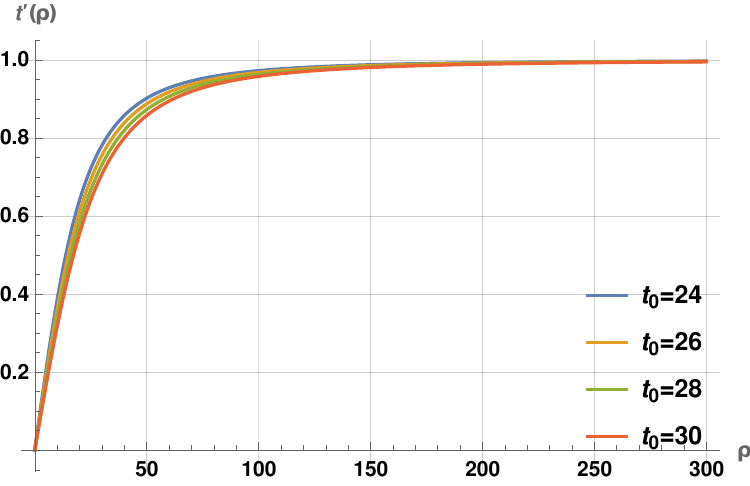}
  \end{subfigure}%
  \begin{subfigure}
    \centering
    \includegraphics[width=.5\linewidth]{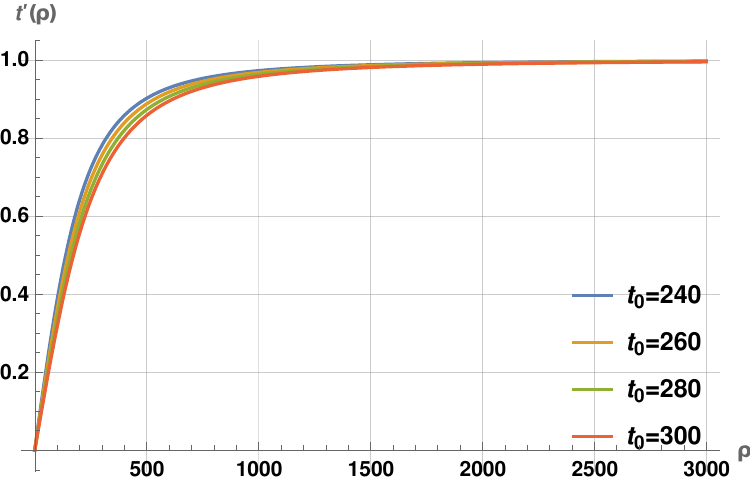}
  \end{subfigure}
 \caption{Variation of volume complexity surfaces $t(\rho)$ and their derivative $t'(\rho)$ with radial distance $\rho$ for 4-dim Lifshitz Kasner cosmology.}
 \label{fig:complexity_surface_radial_corrd_lifshitz_four_dim_numerical-ii}
\end{figure}

As in the earlier cases, it is instructive to study the complexity surface
equation numerically, so we solve
(\ref{lifshitz_metric_z=2,d=2, theta=0_volume functional_eom_t_rho_coords})
with boundary conditions for $t(\rho)$ and $t'(\rho)$ at the boundary
$\rho=\epsilon_\rho$ with appropriate numerical values for $\epsilon_\rho$, along
similar lines as described earlier. The numerical results are shown in
Fig.~\ref{fig:complexity_surface_radial_corrd_lifshitz_four_dim_numerical-ii}
which display the variation of the complexity surfaces and their
derivatives with $\rho$ for various $t_0$ values. The plots show that
the transition to the lightlike regime is slower with nontrivial
Lifshitz $z$-exponent (relative to $AdS$, $z=1$).

Perturbatively we can compute holographic complexity in the regime
$\rho \lesssim t_0$ where $t'(\rho)<1$, approximating complexity
(\ref{lifshitz_metric_z=2,d=2, theta=0_volume functional}) as
\begin{equation} \label{lifshitz_metric_z=2,d=2, theta=0_volume functional-i}
C \sim \frac{V_2 R^2}{G_4} \int_{\epsilon_\rho}^{\rho_\Lambda} d\rho \Biggl[\frac{t(\rho)}{\rho^2}\left({1-\frac{t'(\rho)^2}{2}}\right)\Biggr].
\end{equation}
For the perturbative solution
(\ref{lifshitz_metric_z=2,d=2, theta=0_pertubative_sol_t}), holographic
complexity after truncating
(\ref{lifshitz_metric_z=2,d=2, theta=0_volume functional-i}) up to
next-to-leading-order in $t_0$ is given by
\begin{equation} \label{lifshitz_metric_z=2,d=2, theta=0_volume functional-iv}
C \approx \frac{V_2 R^2}{G_4} \Biggl[t_0 \left(\frac{1}{\epsilon_\rho
   }-\frac{1}{\rho_\Lambda}\right)+\frac{\left(\rho_\Lambda^3-\epsilon_\rho ^3\right)}{24 t_0^3}+{\cal O}\left(\frac{1}{t_0}\right)^5\Biggr].
\end{equation}
We have described the $z=2$ Lifshitz Kasner cosmology so far: other
$z=d_i$ cases in $(d_i+2)$-dims exhibit similar behaviour. The
($t,\rho$) coordinates with $\rho\sim r^z$ in Lifshitz Kasner
cosmologies allow us to conveniently see that the complexity surfaces
become lightlike in the bulk. The metric in (\ref{LifCosExp}) is
recast as
\be\label{genLif-C}
ds^2 = {R^2}\left(-\frac{dt^2}{\rho^2}+\frac{d\rho^2}{\rho^2}+
  {t^{2/z}\over\rho^{2/d_i}} dx_i^2\right)\quad\ra\quad
C = {V_{d_i}\,R^{d_i}\over G_{d_i+2}}\int d\rho\,{t(\rho)\over\rho^2}\sqrt{1-(t'(\rho))^2}\,.
\ee
The leading divergence in $C$ is
\begin{equation}
\label{C-Lif4cosmo}
C \sim \frac{R^{d_i}}{G_{d_i+2}}\,{V_{d_i}\over \epsilon_r^{d_i}}\,t_0\,,
\qquad\qquad
\epsilon_\rho = \epsilon_r^z = \epsilon_r^{d_i}\,.
\end{equation}
The above equation shows the linear time growth of complexity in 
Lifshitz Kasner. Further, from (\ref{C-Lif4cosmo}), we can show that
\begin{equation}
\label{CG-Lif-cosmo}
\frac{dC}{dt_0}\ \sim\   
N_{dof}\, V_{d_i}\Lambda_{_{UV}}^{d_i}\,,
\end{equation}
where $\epsilon_\rho\sim\epsilon_r^{d_i}\equiv\Lambda_{_{UV}}^{d_i}$.
\begin{figure}
 \centering
    \includegraphics[width=0.5\linewidth]{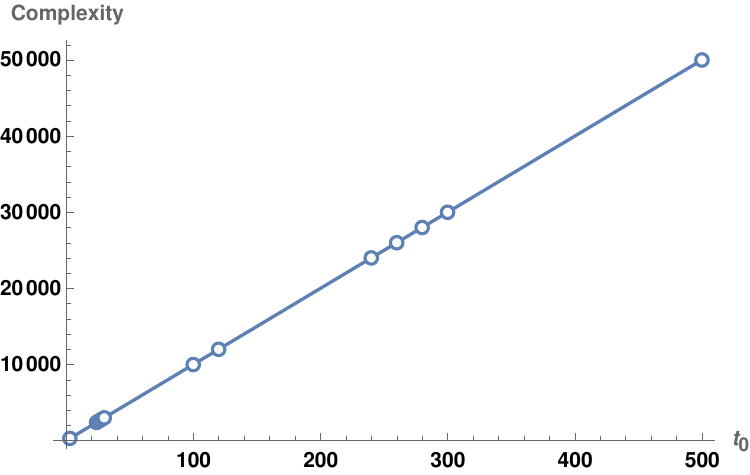}
    \caption{Variation of complexity with $t_0$ for 4-dim Lifshitz
      Kasner cosmology.}
    \label{figurecomplexityt0lifshitzz2d2}
\end{figure}
The observations in
sec.~\ref{sec:lightliket(r)} and sec.~\ref{NCADSK} thus apply here as
well upon analysing (\ref{genLif-C}), and the dual state appears to
have vanishingly low complexity as one approaches the singularity.
This is vindicated in Fig.~\ref{figurecomplexityt0lifshitzz2d2} which
shows holographic complexity plotted against $t_0$ which reveals a
linear decrease as the anchoring time slice approaches the singularity,
\ie\ $t_0\ra 0$.

\section{Holographic entanglement entropy:\ $AdS$ Kasner etc}
\label{EE-cosmologies}

We will review the discussion in \cite{Manu:2020tty} here.
Classical extremal surfaces in cosmological backgrounds are parametrized
by $\left(t(r),x(r)\right)$, $\Delta x =l$, and $t(r) \xrightarrow{r \rightarrow 0} t_0$. The time function $t(r)$ exhibits nontrivial bending due to the
time-dependence. This extremal surface is located at a constant $t$
slice on the boundary denoted by $t=t_0$ and dips into the bulk up to
the turning point and returns to $t_0$. Here $l$ is the width of the
strip along the $x$ direction\ (taking some $x^1=x$), and the extremal
surface wraps the other $x_{j \neq 1}$ directions. The holographic
entanglement entropy becomes
\be
\label{SEE-CES}
S_{\rm EE} = \frac{1}{4 G_{d_i+2}}\int \prod_{x_j \neq x}^{j=1,2,...,(d_i-1)}\left(\phi^{1/d_i}dx_j\right) \sqrt{\frac{e^f}{\phi^{(d_i-1)/d_i}}(-dt^2+dr^2)+\phi^{2/d_i}dx^2}\,.
\ee
The absence of $x(r)$ in (\ref{SEE-CES}) leads to a conserved conjugate
momentum: solving for $x'(r)$ gives
\be\label{x'(r)}
(x'(r))^2=A^2\,\frac{\frac{e^f}{\phi^{(d_i+1)/d_i}}
  \left(1-t'(r)^2\right)}{\phi^2-A^2}\,.
\ee
Substituting back into (\ref{SEE-CES}) gives
\be\label{SEE-CES-i}
S_{\rm EE} =\frac{V_{d_i-1}}{4 G_{d_i+2}} \int dr \left(\frac{e^{f/2} \phi^{(3-1/d_i)/2}}{\sqrt{\phi^2-A^2}} \right)\sqrt{1-t'(r)^2} .
\ee
At the turning point, $x'(r) \rightarrow \infty$ implying $A=\phi_*=\frac{t_*}{r_*^{|m|}}$ (in the approximation $\phi$ is nonvanishing and $t'(r)\ll 1$) where $t_*=t(r_*)$ (since, $\phi =t^k r^m$, $k=1$, $m=-|m|<0$).

For the $AdS$ Kasner spacetime, (\ref{x'(r)}), (\ref{SEE-CES-i}), simplify
to
\begin{eqnarray}\label{EE-AdSK}
x'(r)^2=A^2\left(\frac{1}{t^{2/d_i}}\right)\frac{1-t'(r)^2}{\frac{t^2}{r^{2 d_i}}-A^2}\,, \qquad S_{\rm EE} =\frac{V_{d_i-1}}{4 G_{d_i+2}} \int dr \left(\frac{t^{2-1/d_i}}{r^{2 d_i}} \right)\frac{\sqrt{1-t'(r)^2}} {\sqrt{\frac{t^2}{r^{2 d_i}}-A^2}}\,. 
\end{eqnarray}
Using $A=\phi_*$, $u=\frac{r}{r_*}$, we find the width scaling
\begin{eqnarray}\label{l-scaling}
\frac{l}{2}=\int_0^{r_*} dr x'(r) = r_* \int_0^1 \frac{du}{t^{1/d_i}}\frac{\sqrt{1-t'(r)^2}}{\sqrt{(\phi/\phi_*)^2-1}}\ \ \Rightarrow\ \ 
l \sim r_*\,, \quad A=\phi_*=\frac{t_*}{r_*^{d_i}}\,.
\end{eqnarray}
For a subregion anchored at a time slice $t_0\gg 0$ far from the
singularity, the RT/HRT surface bends in time mildly away from the
singularity. The turning point is $(t_*,r_*)$, with $A>0$ as above for
finite size subregions. The IR limit where the subregion becomes the
entire space is defined as\ $l\ra\infty$ and we expect $r_*\ra\infty$
so the surface extends deep into the interior: here $A=0$. In the
semiclassical region far from the singularity $t_0\ra\infty$, solving
the extremization equation perturbatively for
$t(r)=t_0+\sum_{n \in \mathbb{Z}_+} c_nr^n$
shows that $t_*\gtrsim t_0$ with $t'(r)\ll 1$ for finite subregions
\cite{Manu:2020tty}\ (reviewed numerically in App.~\ref{FA-EE-AdS5K}).
This perturbative analysis is similar to that for the complexity
surface discussed earlier (see Fig.~\ref{t[r]-plots}).
Analysing this in the IR limit is more tricky. In what follows we
will analyse this numerically for the entangling RT/HRT surfaces and
find results similar to those for the complexity surfaces.

\section{Entanglement, $AdS$-Kasner: numerical results}
\label{EE-AdSK-sec}

In this section, we will numerically analyze the codim-2 RT/HRT
surfaces for entanglement entropy numerically, building on the studies
in \cite{Manu:2020tty}, along the same lines as complexity. First we
obtain the perturbative solution of the equation of motion for $t(r)$ and
then use this perturbative solution for boundary conditions to solve
numerically. The extremization equation for $t(r)$ following from the
entanglement area functional (\ref{EE-AdSK}) is\ (there is a
typo in one of the corresponding equations in \cite{Manu:2020tty} but
the analysis there is correct)
\begin{eqnarray}
\label{EOM-t[r]-EE}
\left(1-t'(r)^2\right)\left(d_i^2 t'(r)+\frac{r\left(t(r)^2-A^2 r^{2 d_i}\right)}{t(r)^3}-\frac{d_i r}{t(r)}\right)-\frac{\left(t(r)^2-A^2 r^{2 d_i}\right)d_i r t''(r)}{t(r)^2}=0\,.
\end{eqnarray}
We will focus on solving (\ref{EOM-t[r]-EE}) in the IR limit $A=0$
(\ref{l-scaling}) for infinitely wide strip subregions in
$AdS_{5,7}$-Kasner spacetimes in sec.~\ref{EE-AdS5K-subsec} and
sec.~\ref{EE-AdS7K-subsec} respectively.
The above equation in the IR limit $A=0$ becomes
\be\label{EOM-t[r]-EE-IR}
d_i\,r\,t(r)\,t''(r)\, -\,
\left(1-t'(r)^2\right)\big(d_i^2\,t(r)\,t'(r)-(d_i-1)r \big) = 0\,.
\ee

\subsection{Holographic entanglement entropy in $AdS_5$-Kasner}
\label{EE-AdS5K-subsec}
\begin{figure}[h]
\begin{subfigure}
  \centering
  \includegraphics[width=.5\linewidth]{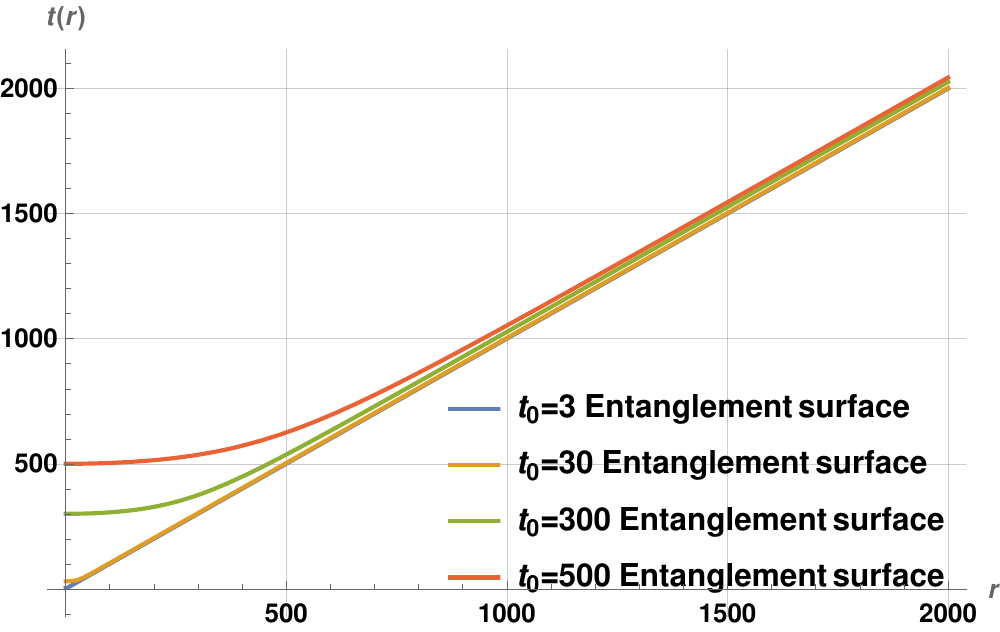}
\end{subfigure}
\begin{subfigure}
  \centering
  \includegraphics[width=.5\linewidth]{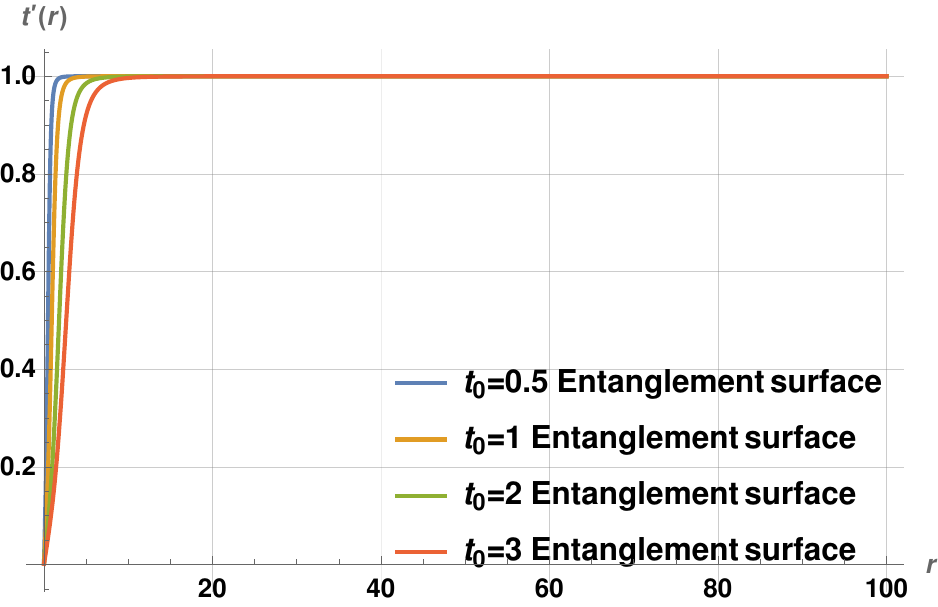}
\end{subfigure}
\begin{subfigure}
  \centering
  \includegraphics[width=.5\linewidth]{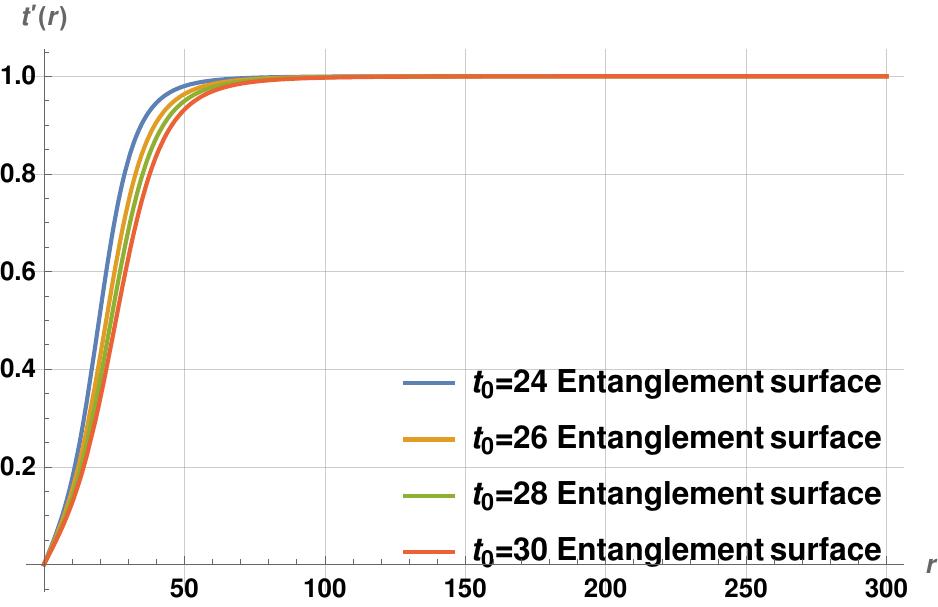}
\end{subfigure}
\begin{subfigure}
  \centering
  \includegraphics[width=.5\linewidth]{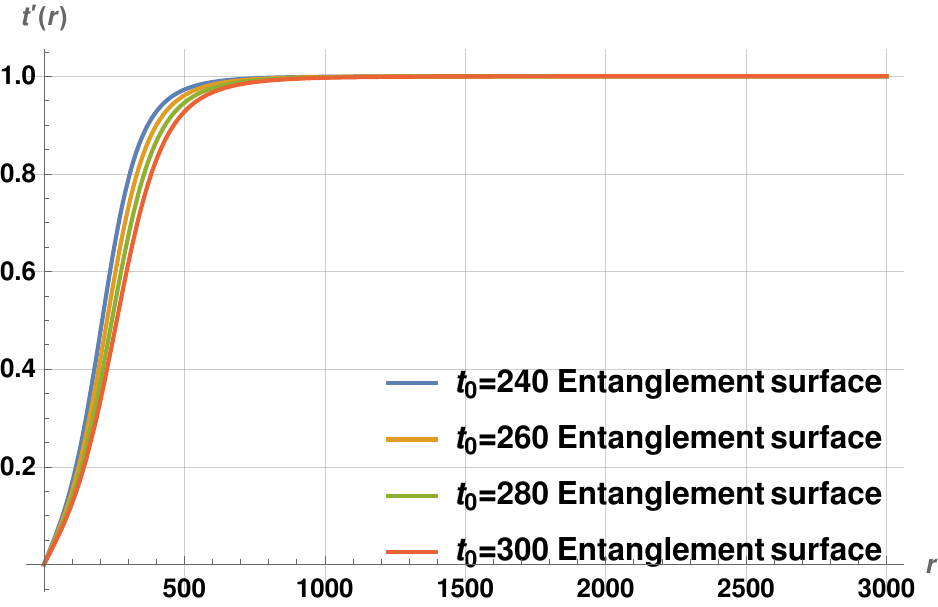}
\end{subfigure}
\caption{Variations of the RT/HRT surface $t(r)$ vs $r$ and $t'(r)$
  vs $r$ in $AdS_5$-Kasner spacetime for various $t_0$ slices.}
\label{t[r]-plots-AdS5K-EE}
\end{figure}
For $AdS_5$-Kasner with $d_i=3$, the IR limit (\ref{EOM-t[r]-EE-IR}) becomes
\begin{eqnarray}
\label{EOM-t(r)-AdS5K-EE}
3\,r\,t(r)\,t''(r)\, -\,
\left(1-t'(r)^2\right)\big(9\,t(r)\,t'(r)-2r \big) = 0\,.
\end{eqnarray}
The perturbative solution of (\ref{EOM-t(r)-AdS5K-EE}) using the
ansatz $t(r)=t_0+\sum_{n \in \mathbb{Z}_+}c_n r^n$, after truncating, is
\begin{eqnarray}
\label{soln-t(r)-AdS5K-EE}
t(r)\sim t_0+\frac{r^2}{6 t_0}\,.
\end{eqnarray}
The numerical solutions of (\ref{EOM-t(r)-AdS5K-EE}) and their
derivatives are shown in Fig.~\ref{t[r]-plots-AdS5K-EE}. 
This shows that the behaviour of RT/HRT surfaces is similar to 
complexity surfaces, as discussed earlier. In particular, the RT/HRT
surface for lower $t_0$ (closer to the singularity) becomes lightlike
earlier in comparison to RT surfaces with higher $t_0$ values. Thus as
we approach the singularity with $t_0\ra 0$, entanglement entropy
becomes vanishingly small. In particular near the singularity,
entanglement entropy vanishes as did complexity. There is an extreme
thinning of the degrees of freedom near the singularity.

For $AdS_4$-Kasner, the results are qualitatively similar but the
numerics turn out to not be as clean for just technical rather than
physics reasons so we do not discuss this.

\subsection{Holographic entanglement entropy, $AdS_7$-Kasner}
\label{EE-AdS7K-subsec}

The $t(r)$ equation of motion in the IR limit (\ref{EOM-t[r]-EE-IR})
for $AdS_7$-Kasner spacetime with $d_i=5$ is:
\begin{eqnarray}
\label{EOM-t(r)-AdS7K-EE}
5\,r\,t(r)\,t''(r)\, -\,
\left(1-t'(r)^2\right)\big(25\,t(r)\,t'(r)-4r \big) = 0\,.
\end{eqnarray}
The perturbative solution of (\ref{EOM-t(r)-AdS7K-EE}) using the
ansatz $t(r)=t_0+\sum_{n \in \mathbb{Z}_+}c_n r^n$ is obtained as:
\begin{eqnarray}
\label{soln-t(r)-AdS7K-EE}
t(r)=t_0+\frac{r^2}{10 t_0}-\frac{9 r^4}{1000 t_0^3}.
\end{eqnarray}
\begin{figure}[h]
\begin{subfigure}
  \centering
  \includegraphics[width=.5\linewidth]{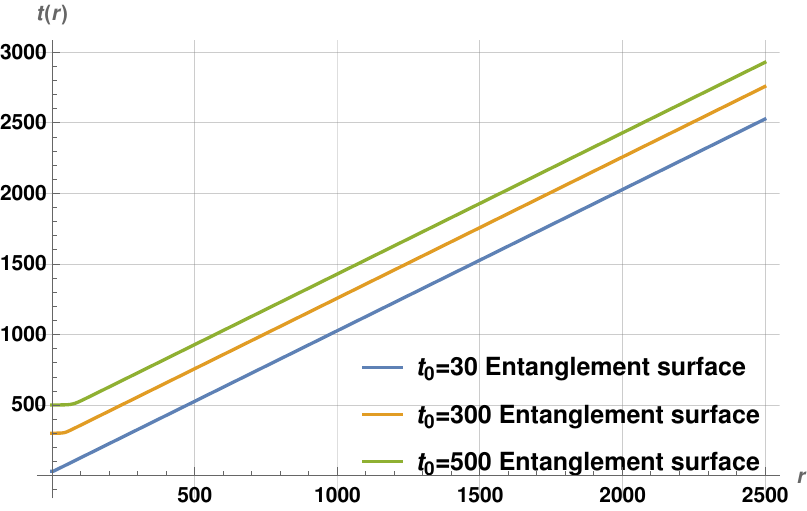}
\end{subfigure}
\begin{subfigure}
  \centering
  \includegraphics[width=.5\linewidth]{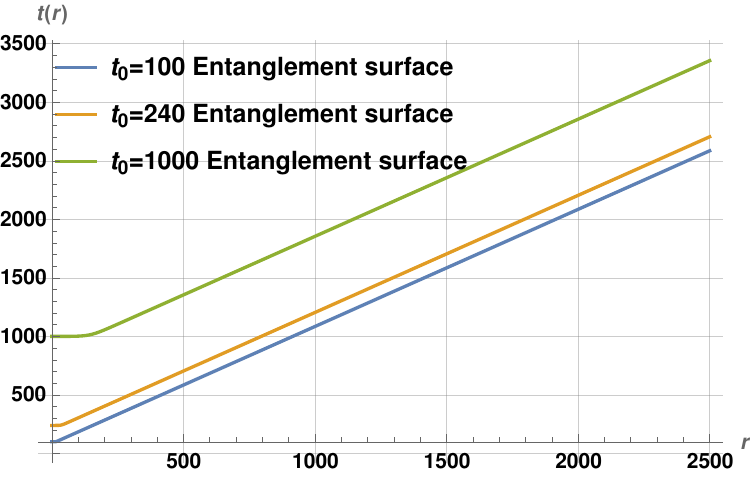}
\end{subfigure}%
\begin{subfigure}
  \centering
  \includegraphics[width=.5\linewidth]{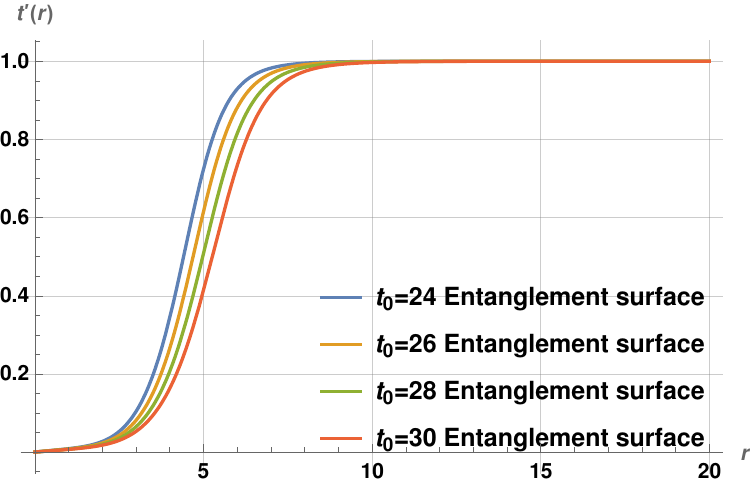}
\end{subfigure}
\begin{subfigure}
  \centering
  \includegraphics[width=.5\linewidth]{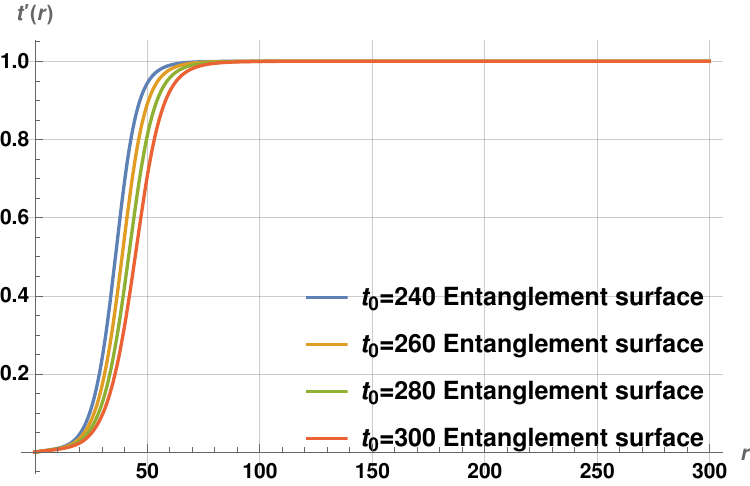}
\end{subfigure}
\caption{Numerical plots of the RT/HRT surfaces $t(r)$ versus $r$ and
  $t'(r)$ versus $r$ in AdS$_7$-Kasner spacetime for different $t_0$ slices.}
\label{LL-i-t'[r]-EE-AdS7K}
\end{figure}
The numerical solutions of (\ref{EOM-t(r)-AdS7K-EE}) and their
derivatives are shown in Fig.~\ref{LL-i-t'[r]-EE-AdS7K}: we find
the entangling surfaces $t(r)$ have qualitatively similar behaviour 
as in $AdS_5$-Kasner.

\subsection{Numerical computation of holographic entanglement entropy}
\label{NC-HEE-AdSKasner}

\begin{figure}[h]
\begin{subfigure}
  \centering
  \includegraphics[width=.5\linewidth]{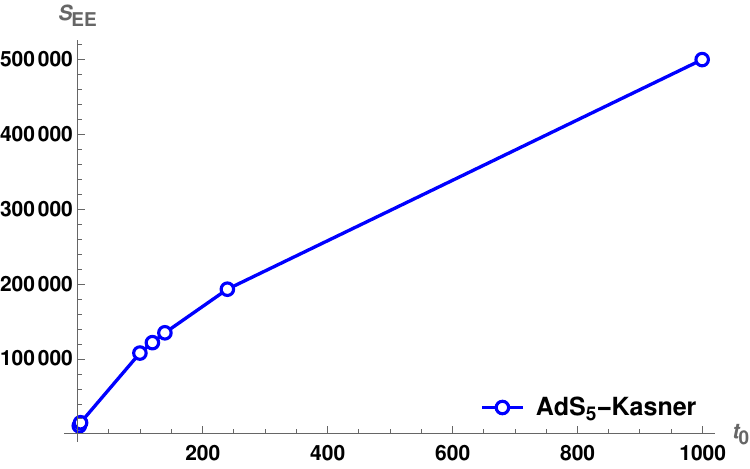}
\end{subfigure}
\begin{subfigure}
  \centering
  \includegraphics[width=.5\linewidth]{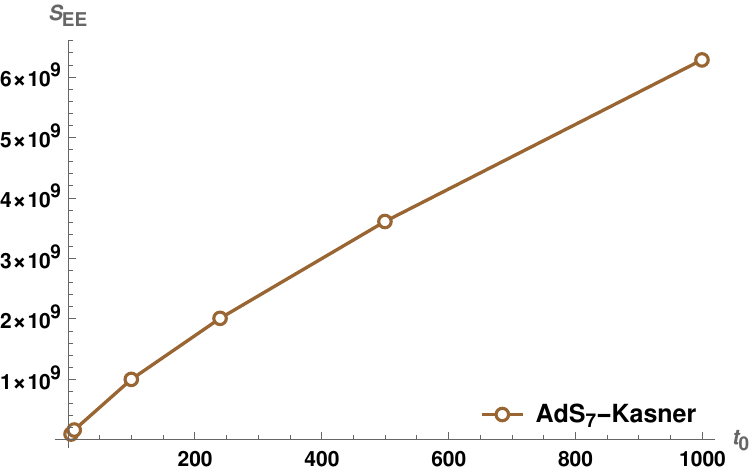}
\end{subfigure}
\caption{Numerical plots of holographic entanglement entropy versus $t_0$
  for $AdS_{5,7}$-Kasner.}
\label{SEE-AdSK-N}
\end{figure}

The holographic entanglement entropy (\ref{EE-AdSK}) in the IR limit
$A=0$ for $AdS_{d_i+2}$-Kasner spacetime becomes
\begin{eqnarray}
\label{EE-AdSK-summary}
S_{\rm EE}=\frac{R^{d_i}\,V_{d_i-1}}{4 G_{d_i+2}}\int_\epsilon dr
\left(\frac{t(r)^{(d_i-1)/d_i}}{r^{d_i}}\right)\sqrt{1-(t'(r)^2)}\,.
\end{eqnarray}
We evaluated the integrals appearing in (\ref{EE-AdSK-summary})
numerically for $AdS_{5,7}$-Kasner spacetimes by taking
$\epsilon=0.01$ (and setting the lengthscales $R, V_{d_i-1}, G_{d_i+2}$
to unity). 
As an order-of-magnitude estimate with $t_0\sim 1000$, we then have\
$S_{EE}\sim {t_0^{2/3}\over\epsilon^2}\sim 10^6$.\
The variation of entanglement entropy in the IR limit in
$AdS_{5,7}$-Kasner with $t_0$ is shown in Fig.~\ref{SEE-AdSK-N}.
This indicates that entanglement entropy in AdS-Kasner spacetime
decreases as $t_0$ decreases, and eventually becomes zero when
$t_0\ra 0$. This is also consistent with what we observed in
Fig. \ref{t[r]-plots-AdS5K-EE} where we see that the entangling
surfaces $t(r)$ become lightlike earlier for anchoring time slice
$t_0$ closer to the singularity.

We now compare the IR entangling RT/HRT surfaces for
$AdS_5$-Kasner (Fig.~\ref{t[r]-plots-AdS5K-EE}) and
$AdS_7$-Kasner (Fig.~\ref{LL-i-t'[r]-EE-AdS7K}), as we had done for 
complexity surfaces in sec.~\ref{NCADSK}. 
We see that the surfaces approach the lightlike regime earlier in
$AdS_7$-Kasner relative to $AdS_5$-Kasner, similar to the complexity
surfaces. Here again this appears to stem from the amplification
factor ${1\over r^{d_i-1}}$ of the lightlike factor $\sqrt{1-(t')^2}$
in entanglement entropy, so the effective thinning of the degrees
of freedom is more rapid in higher dimensions.

As a mathematical observation, by comparing the $t(r)$ plots in
$AdS$-Kasner, we see that the complexity surfaces become lightlike
earlier than the entangling RT/HRT surfaces. This can be seen to
follow from the equations of motion which are similar in structure
but differ in the numerical factors that appear: for $AdS_5$-Kasner,
we have from (\ref{EOM-t(r)-AdS5K-EE}), (\ref{EOM-t[r]-AdS5-Kasner}),
\be\label{EE-C}
1-(t')^2 = {{3\over 2} r\,t\,t''\over {9\over 2} t\,t'-r}\ \ \ [{\rm EE}]\,;
\qquad 1-(t')^2 = {r\,t\,t''\over 4 t\,t'-r}\ \ \ [{\rm complexity}]\,.
\ee
the denominator factors are comparable so the relative factor of
${3\over 2}$ makes $(1-(t')^2)$ larger so $t'(r)$ is smaller for the
entangling surface $t(r)$.

Explicit expressions for holographic entanglement entropy
(\ref{SEE-CES-i}) can be obtained for hyperscaling violating cosmologies
using (\ref{HDBC-ntheta}). The exponents for $t(r)$ are fairly
nontrivial. The $r$-scalings give the leading divergence as\
$S_{EE}\sim {V_{d_i-1}\over G_{d_i+2}}\,{R^{d_i-\theta}\over\epsilon^{d_i-\theta-1}}$\
where we have reinstated the dimensionful bulk scale $R$ (which
can be done simply on dimensional grounds).\ This can be recast
as
\be\label{EEscalings-hv}
S_{EE}\sim {V_{d_i-1}\over G_{d_i+2}}\,{R^{d_i-\theta}\over\epsilon^{d_i-\theta-1}}
\ \sim\ N_{eff}(\epsilon)\,{V_{d_i-1}\over\epsilon^{d_i-1}}\,,\qquad
N_{eff}(\epsilon)={R^{d_i-\theta}\over G_{d_i+2}}\,\epsilon^\theta\,,
\ee
where $N_{eff}$ is the effective scale-dependent number of degrees
of freedom evaluated at the UV cutoff length $\epsilon$\
(see \cite{Ryu:2006ef}, \cite{Barbon:2008ut}). In concrete gauge/string realizations of
hyperscaling violating theories obtained by dimensional reduction of
nonconformal $Dp$-branes, it can be seen that the lengthscales in the
$Dp$-brane description reorganize themselves as the above and 
also match various expectations, including from
considerations of the holographic c-function from a 2-dim dilaton
gravity point of view \cite{Kolekar:2018chf}. For instance, 
the $d_i=2,\ \theta=-{1\over 3}$ case corresponding to the D2-brane
supergravity phase with $G_4\sim {G_{10}\over R^6}$ after the transverse
sphere reduction gives $N_{eff}=N^2\,(g_{YM}^2N\epsilon)^{-1/3}$ which
ends up being consistent with the regime of validity of the
D2-supergravity phase. By comparison the complexity scalings then
are less obvious. The leading divergence of complexity in hyperscaling
violating theories can be expressed as
\be
C\ \sim\ {V_{d_i}\over G_{d_i+2}}\,
\Big({R\over\epsilon}\Big)^{d_i-\theta-{\theta\over d_i}}\
\sim\ N_{eff}(\epsilon)\,{V_{d_i}\over\epsilon^{d_i}}\,
\Big({R\over\epsilon}\Big)^{\!\!\!^{-{\theta\over d_i}}}\,,
\ee
using $N_{eff}$ in (\ref{EEscalings-hv}).
The extra factor $({R\over\epsilon})^{{-\theta\over d_i}}$ arising from
the extra metric factor for codim-1 surfaces (relative to codim-2)
cannot be obviously recast in terms of field theory parameters
once $N_{eff}$ is pulled out\ (see
\cite{Swingle:2017zcd,Alishahiha:2018tep,Zhu:2020zti,Omidi:2022whq}
for other complexity studies). Of course this can be expressed in
terms of some effective UV cutoff\
${\tilde\Lambda}^{d_i-{\theta\over d_i}}\,R^{-{\theta\over d_i}}$.
The scalings of complexity with time are also nontrivial. It
would be interesting to understand this better.

The numerical plots of the entangling RT/HRT surfaces for the hyperscaling
violating cosmology with $d_i=4, \ \theta=-1$ are qualitatively
similar to the above for sufficiently high $t_0$: away from this,
there appear to be some numerical issues\ (as well as for the
$d_i=2,\ \theta=-{1\over 3}$ case), similar to the $AdS_4$ Kasner
case stated earlier. So we do not discuss these in detail.

\section{Discussion}\label{summary}

We have studied holographic volume complexity and entanglement entropy
in various families of cosmologies with Big-Bang/Crunch singularities,
some of which were studied previously in \eg\
\cite{Das:2006dz}-\cite{Engelhardt:2016kqb}. These include $AdS$,
hyperscaling violating and Lifshitz asymptotics.  Focussing on
isotropic Kasner-like singularities, we saw that higher dimensional
complexity and entanglement can be recast in terms of that in
2-dimensional dilaton gravity theories obtained by dimensional
reduction \cite{Bhattacharya:2020qil}, and the resulting expressions
are compactly written in terms of entirely 2-dim variables.

\begin{figure}[h] 
\includegraphics[width=12pc]{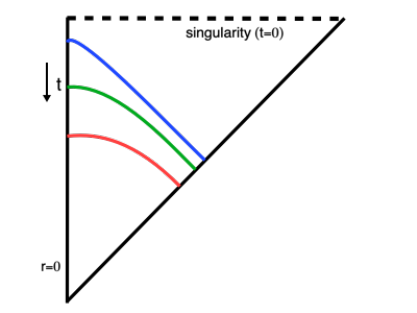}
\begin{minipage}[b]{26.5pc}
\caption{{ \label{figbbCcmpxtySurf}
\footnotesize{
  Cartoon of complexity and IR entanglement surfaces at various anchoring
  time slices $t$ on the boundary ($r=0$) in holographic cosmologies with
  Kasner-like Big-Crunch singularities. The extremal surfaces bend
  away from the singularity (dotted line, $t=0$) and approach
  lightlike regimes eventually (approaching faster as
  $t\ra 0$).\newline
}}}
\end{minipage}
\end{figure}  
The equation of motion for the complexity and IR entangling surfaces
obtained by extremization of the complexity and entanglement
functionals can be solved perturbatively near the holographic
boundary: using this allows us to extract boundary conditions for
numerical solutions of the surfaces.
In the numerics, we impose a near-boundary cutoff $\epsilon$:
the interior end approaches a lightlike regime so no interior
regulator is required. The complexity plots appear in
Figs.~\ref{LL-i-t[r]}-\ref{LL-i}, \ref{LL-i-AdS4K-Num},
\ref{LL-i-AdS7K-Num} for $AdS_{5,4,7}$ Kasner,\ Figs.~\ref{LL-i-HSV},
\ref{LL-i-HSV-theta=-1} for hyperscaling violating cosmologies, and
Fig.~\ref{fig:complexity_surface_radial_corrd_lifshitz_four_dim_numerical-ii}
for Lifshitz Kasner, and those of entanglement appear in
Fig.~\ref{t[r]-plots-AdS5K-EE}, Fig.~\ref{LL-i-t'[r]-EE-AdS7K} for
$AdS_{5,7}$ Kasner. 

Overall this shows that the surfaces begin spacelike near the
boundary, bend in the direction away from the location of the
singularity and transition to lightlike in the interior
(sec.~\ref{sec:lightliket(r)}).  For instance in (\ref{EE-C}),
(i) with $(t')^2\ll 1$ (spacelike), we see by using a series
expansion for $t(r)$ that $t'(r)>0$, and (ii) with $1-(t')^2\sim 0^+$
(lightlike), we see that $t''>0$. As the anchoring time slice
is moved towards the singularity, the spacelike part shrinks and the
transition to lightlike is more rapid.\ The overall picture depicting
a future Big-Crunch singularity is shown in Fig.~\ref{figbbCcmpxtySurf}
above (which is top-bottom reflected relative to the plots):
note that $t\equiv |t|$ here so our analysis applies equally well
to past Big-Bang singularities (\eg\ in (\ref{EE-C}), $t\ra -t$ is
a symmetry).
 The complexity and entanglement functionals
contain a $\sqrt{1-(t')^2}$ factor so that the lightlike regimes give
vanishing contributions:\ see Figs.~\ref{NC-AdSK}, \ref{NC-hvLif-g},
\ref{figurecomplexityt0lifshitzz2d2} (complexity) and
Fig.~\ref{SEE-AdSK-N} (EE). Thus the near singularity region has
vanishingly low complexity and entanglement and the ``dual Kasner
state'' in all these theories corresponds to the effective number of
qubits being vanishingly low, consistent with spatial volumes
undergoing a Crunch.  Our results corroborate those in
\cite{Barbon:2015ria} for volume complexity, and in
\cite{Caputa:2021pad} from holographic path integral optimization, in
$AdS$ Kasner. However our analysis (in particular numerically) is more
detailed and applies to various families of cosmologies which are in
the same ``universality class'' in the scaling behaviour
(\ref{universality}) near the singularity.  Our entanglement analysis
develops further the semiclassical perturbative study in
\cite{Manu:2020tty}, where the entangling surfaces were shown to bend
away from the singularity (and quantum extremal surfaces are driven
far away). Our numerics is consistent with the behaviour of entangling
surfaces for finite subregions which only bend mildly
(App.~\ref{FA-EE-AdS5K}).

It is worth noting that in the region very near the singularity, where
the transition to lightlike is rapid, the anchoring time slice $t_0$
eventually becomes comparable to the cutoff $\epsilon$,
\ie\ $t_0\sim\epsilon$.  At this point, it is perhaps best to say that
the semiclassical gravity framework here becomes unreliable: so in
concluding complexity to be vanishingly low as $t_0\ra 0$, we are
extrapolating the decreasing complexity to the very near singularity
region. While this appears reasonable, it is worth understanding the
very near singularity region more elaborately.  At a very basic level,
our analysis of holographic volume complexity and entanglement
(building on those in \cite{Manu:2020tty,Goswami:2021ksw}) and those
in \cite{Barbon:2015ria}, \cite{Caputa:2021pad}\ (as well as the
limiting surface in the black hole interior \cite{Hartman:2013qma})
suggests that these sorts of spacelike Kasner cosmological
singularities are excluded from the entanglement wedge of observers,
as defined by the extremal surfaces that self-consistently steer clear
of the vicinity of the singularity. This is a kind of
{\bf{``entanglement wedge cosmic censorship''}}\ (we thank Sumit Das
for this phrase!). In some sense, this is reassuring since if it were
not true, it would amount to a breakdown of the semiclassical gravity
framework here, and thereby inconsistencies. Perhaps studying null
singularities will reveal qualitatively new behaviour in light of the
studies of the holographic duals in \cite{Das:2006pw} and quantum
extremal surfaces in \cite{Goswami:2021ksw} which bend towards the
singularity.\\
(The extremal surfaces bending away from the singularity 
is reminiscent of the absence of spacelike surfaces anchored
at the future boundary in de Sitter space \cite{Narayan:2015vda}:
perhaps more general structures \cite{Doi:2022iyj,Narayan:2022afv}
may be of value as near singularity probes.)

We have focussed on the isotropic Kasner subfamily which is natural
from the point of view of reduction to 2-dimensions. However it
is likely that there are more general spacetimes with hyperscaling
violating and Lifshitz asymptotics with general anisotropic Kasner
singularities, analogous to the general anisotropic Kasner
spacetimes in $AdS$. The constraint $\sum_ip_i=1$ would then imply
that holographic volume complexity would be the same as in our
analysis, although entanglement entropy, requiring a specification
of the boundary spatial subregion would depend on the spatial
orientation of the subregion.

More general $AdS$-BKL-type singularities were also studied in
\cite{Awad:2008jf}. In these cases, spatial curvatures force BKL
oscillations between various Kasner regimes (starting with some Kasner
exponent negative), which continue indefinitely in the absence of
external scalars \cite{Landau,BKL,BK}. In the presence of the scalar
$\Psi$ as we have, the BKL oscillations lead to attractor-type basins
eventually (with all Kasner exponents positive). Holographic
entanglement requires defining a spatial subregion and thus would
appear to evolve along BKL oscillations. Since the volume complexity
functional for anisotropic $AdS$-Kasner backgrounds is similar, the
evolution of complexity naively appears insensitive to these BKL
oscillations, but it would be interesting to explore complexity more
carefully to see the role of spatial curvatures.

The effectively 2-dim nature of our bulk analysis suggests effective
dual 1-dim qubit models governing complexity. In $AdS$ and Lifshitz
Kasner, the complexity decrease with time is linear whereas in
hyperscaling violating theories, it is not.  These latter
theories with nonzero $\theta$ have effective spatial dimensions
$d_{eff}=d_i-\theta$: it might be interesting to study
effective 1-dim qubit models simulating this (recalling
the general arguments in \cite{Susskind:2014rva}).

Finally it is worth noting that the Kasner singularities we have been
discussing have time dependence that does not switch off
asymptotically. This reflects in the nontrivial Kasner scale $t_K$
lingering in our expressions: for instance (\ref{C-t0}) after
reinstating $t_K$ is really\
$C\sim N_{dof}\, V_{d_i}\Lambda_{_{UV}}^{d_i}\, {t_0\over t_K}$\
so this perhaps cannot be extrapolated to asymptotically large
timescales $t_0\gg t_K$.  The main merit of these models is the
simplicity of the bulk in the vicinity of the singularity.  Perhaps,
as in \cite{Goswami:2021ksw} for quantum extremal surfaces, asymptotic
regions with no time-dependence can be appended beyond $t_0>t_K$ with
appropriate boundary conditions. In this case, the extremal surfaces
becoming lightlike hitting the past horizon here
(Fig.~\ref{figbbCcmpxtySurf}) must instead presumably be extended to
these asymptotic far-regions\ (translating the question of the
behaviour at the past horizon to the behaviour asymptotically).
This hopefully will lead to better understanding of the (non-generic)
initial conditions in the asymptotic regions that give rise to this
``dual Kasner state'' and its low complexity.

\vspace{7mm}

{\footnotesize \noindent {\bf Acknowledgements:}\ \ We (especially KN)
  are particularly grateful to Sumit Das for very insightful early
  discusssions on low holographic complexity in the vicinity of
  cosmological singularities.  We also thank Pawel Caputa, Abhijit
  Gadde, Alok Laddha, A. Manu and Rob Myers for useful discussions,
  and Sumit Das and Rob Myers for useful comments on a draft.
  GY would also like to thank Krishna Jalan, Pankaj Saini, Harsh Rana
  and Ashutosh Singh for helpful discussions about numerical
  calculations. GY would like to thank the Isaac Newton Institute for
  Mathematical Sciences for support and hospitality during the
  programme ``Bridges between holographic quantum information and
  quantum gravity'' while this work was in progress. We thank
  the organizers of the Indian Strings Meeting 2023 (ISM2023), IIT
  Bombay, for hospitality while this work was in progress. This
  work is partially supported by a grant to CMI from the Infosys
  Foundation and by EPSRC Grant Number EP/R0146014/1.}

\appendix

\section{Holographic cosmologies\ $\ra$\ 2-dim}\label{sec:HolCosD}
\setcounter{equation}{0}
\seceqaa
Time-dependent non-normalizable deformations of $AdS/CFT$ were studied
in \cite{Das:2006dz,Das:2006pw,Awad:2007fj,Awad:2008jf} towards
gaining insights via gauge/gravity duality into cosmological (Big-Bang
or -Crunch) singularities. The bulk gravity theory exhibits a cosmological
Big-Crunch (or -Bang) singularity and breaks down while the
holographic dual field theory (in the $AdS_5$ case) subject to a
severe time-dependent gauge coupling $g_{YM}^2=e^\Psi$ (and living on
a time-dependent base space) may be hoped to
provide insight into the dual dynamics: in this case the scalar $\Psi$
controls the gauge/string coupling.  There is a large family of such
backgrounds exhibiting cosmological singularities. Among the simplest
are $AdS$-Kasner theories
\be\label{adsKasner}
ds^2 = {R_{AdS}^2\over r^2} (-dt^2 + \sum_i t^{2p_i} dx_i^2 + dr^2) , \quad
e^\Psi=t^\al\ ;\qquad \sum_i p_i=1\ ,\quad \sum_i p_i^2 = 1 - {1\over 2}\al^2\ .
\ee
For constant scalar $\Psi$ with $\al=0$, the Kasner space is
necessarily anisotropic: the $p_i$ cannot all be equal. In this case,
the gauge theory lives on a time-dependent space but the gauge
coupling is not time-dependent. The isotropic subfamily requires a
nontrivial scalar source $\Psi$ as well. More general backgrounds can
also be found involving $AdS$-FRW and $AdS$-BKL spacetimes
\cite{Awad:2007fj,Awad:2008jf}, all of which have spacelike
singularities. There are also backgrounds with null singularities
\cite{Das:2006pw}. Similar Kasner deformations exist for $AdS_4\times
X^7$ and $AdS_7\times X^4$. For generic spacelike singularities, the
gauge theory response appears singular \cite{Awad:2008jf} while null
singularities appear better behaved \cite{Das:2006pw}. Some of these
spacelike singularities were further investigated in
\cite{Engelhardt:2014mea,Engelhardt:2015gta,Engelhardt:2015gla,
  Engelhardt:2016kqb}

These arise in higher dimensional theories of Einstein gravity with
scalar $\Psi$, a potential $V$, and action
\be\label{actionDdim}
S = {1\over 16 \pi G_D} \int d^Dx \sqrt{-g^{(D)}}\,
\Big( {\cal R} - {1\over 2} (\del\Psi)^2 - V \Big)\ .
\ee
We allow the potential $V$ to also contain metric data, \ie\ it is a
function $V(g,\Psi)$.  Under dimensional reduction with ansatz
(\ref{dimredAnsatz}), we obtain the 2-dim action (\ref{2ddg-action})\
(see the general reviews
\cite{Strominger:1994tn,Grumiller:2002nm,Mertens:2022irh}, of
2-dim dilaton gravity theories and dimensional reduction).
In general these sorts of generic 2-dim dilaton gravity theories
encapsulate various aspects of the higher dimensional gravity
theories, and are perhaps best regarded as effective holographic
models \cite{Narayan:2020pyj}.  These sorts of theories were
considered in \cite{Kolekar:2018chf} towards understanding
holographic c-functions from the 2-dim dilaton gravity point of
view. The 2-dim equations of motion following from (\ref{2ddg-action})
were solved in \cite{Bhattacharya:2020qil} with various families of
asymptotics (flat, $AdS$, hyperscaling violating and Lifshitz) to
obtain various classes of 2-dim cosmologies with Kasner-like
Big-Bang/Crunch singularities.

We now review a little more from \cite{Bhattacharya:2020qil}.
With $AdS$ asymptotics, we have\ $V=2\Lambda$ giving the dilaton
potential in (\ref{2ddg-action}) as $U=2\Lambda\phi^{1/d_i}$\
(\ref{AdSDK-2d}) independent of the scalar $\Psi$.\ Hyperscaling
violating asymptotics\
$ds^2=\frac{{R^{2}}r^{2\theta/d_i}}{r^2} (-dt^2+dr^2+dx_i^2)$\
with nontrivial exponent $\theta$ arise \cite{Dong:2012se} from
dimensional reductions
of nonconformal $Dp$-branes \cite{Itzhaki:1998dd}: after reduction over
the transverse sphere we obtain a $(d_i+2)$-dim action of the form
(\ref{actionDdim}) with $V=2\Lambda\,e^{\gamma\Psi}$, which after
reduction over the $d_i$ spatial dimensions gives (\ref{2ddg-action})
with $U$ in (\ref{HDBC-ntheta}), and the corresponding parameters for
the on-shell backgrounds. Lifshitz asymptotics\
$ds^2 = R^2(-{dt^2\over r^{2z}}+{dr^2\over r^2} + {dx_i^2\over r^2})$\
with nontrivial exponent $z$ requires a further gauge field strength,
which on-shell leads to an action (\ref{actionDdim}) with effective
potential of the form $V=\phi^{-1/d_i}U$ with $U$ in (\ref{LifCosExp}).
Hyperscaling violating Lifshitz theories contain both nontrivial
$z$ and $\theta$ exponents.

Cosmological deformations of the isotropic Kasner kind were found
in \cite{Bhattacharya:2020qil} by solving the 2-dim theories obtained
by reduction over the transverse $d_i$-space. The power law ansatz
(\ref{2d-tr-exp}) for the 2-dim fields\ $\phi,\ e^f,\ e^\Psi$\
describes the vicinity of the singularity. The exponents, fixed
by the 2-dim equations, with various asymptotics are in
(\ref{AdSDK-2d}),\ (\ref{HDBC-ntheta}) and (\ref{LifCosExp}). The
asymptotics are the same as those in the absence of the time-dependence.
For the $AdS$ and hyperscaling violating cases, the solutions for the
$t$- and $r$-parts of the equations of motion end up being compatible
(they are roughly independent).
In general however, the time-dependent backgrounds are more constraining,
particularly in the Lifshitz case where the equations couple the $t$-
and $r$-exponents forcing $\theta=0$ and $z=d_i$.

As in $AdS_5$ Kasner, the scalar $e^\Psi$ controls the gauge coupling
in nonconformal brane theories as well. Taking the exponent $\al>0$
in (\ref{HDBC-ntheta}) amounts to taking the gauge coupling to vanish
at $t=0$ which then leads to diverging $(\dot\Psi)^2\sim {1\over t^\#}$
and thence a bulk singularity.

\section{$g_i, s_i, y_i, v_i$}
\label{appendix-A}
\setcounter{equation}{0}
\seceqbb

\begin{itemize}
\item The iterative solution of (\ref{EOM-t[r]-AdS5-Kasner}) up to $O(r^{30})$ is given as:
{\footnotesize
\begin{eqnarray}
\label{soln-t[r]-30-app}
& & \hskip -0.2in
t(r)=t_0+\frac{r^2}{6
   t_0}-\frac{7 r^4}{216 t_0^3}+\frac{5
   r^6}{3888 t_0^5}-\frac{23 r^8}{31104 t_0^7}+\frac{5671
   r^{10}}{125971200 t_0^9}-\frac{193157 r^{12}}{31744742400 t_0^{11}}-\frac{1451389 r^{14}}{571405363200
   t_0^{13}} \nonumber\\
 & & \hskip -0.2in  +\frac{126271147 r^{16}}{60340406353920
   t_0^{15}}-\frac{171499492421 r^{18}}{211794826302259200
   t_0^{17}}+\frac{2509650528887 r^{20}}{7624613746881331200
   t_0^{19}} \nonumber\\
 & & \hskip -0.2in -\frac{23544237318388621 r^{22}}{213870415600021340160000
   t_0^{21}}+\frac{74037865493904302737
   r^{24}}{2048023099785804353372160000
   t_0^{23}}\nonumber\\
 & & \hskip -0.2in -\,\frac{27221138559698748551
   r^{26}}{2457627719742965224046592000
   t_0^{25}}+\,\frac{2040692465059715118445379
   r^{28}}{640998461863360189735832125440000
  t_0^{27}} \nn\\
 & & \qquad -\frac{414120404436438180460454771
   r^{30}}{480748846397520142301874094080000000
   t_0^{29}}\,.
\end{eqnarray}}
Likewise (\ref{soln-t[r]-ii}), (\ref{soln-t(r)-AdS7K}), (\ref{t[r]-general-soln-NCB-IR-theta=-1by3}), (\ref{t[r]-soln-NCB-IR-theta=-1by3}), (\ref{t[r]-general-soln-NCB-IR-theta=-1}), 
(\ref{t[r]-soln-NCB-IR-theta=-1}), (\ref{lifshitz_metric_z=2,d=2, theta=0_pertubative_sol_t}) display truncated solutions analogous to (\ref{soln-t[r]-30}). The numerical plots do not change much with the truncation.
\item $g_{2,4}$ appearing in (\ref{t[r]-general-soln-NCB-IR-theta=-1by3}) are given as: 
\begin{eqnarray}
\label{ci's-theta=-1by3-general}
& &  g_2=\frac{15-2 \sqrt{2}}{70 t_0}, \ \ \ g_4=\frac{5036 \sqrt{2}-25835}{343000 t_0^3}.
\end{eqnarray}
\item $s_{2,4}$ appearing in (\ref{t[r]-soln-NCB-IR-theta=-1by3}) are given as:
\begin{eqnarray}
\label{ci's-theta=-1by3}
& & s_2=\frac{15-2 \sqrt{2}}{70 t_0}, \ \ \ s_4=\frac{-233-60 \sqrt{2}}{1960 t_0^3}.
\end{eqnarray}
\item $y_{2,4}$ appearing in (\ref{t[r]-general-soln-NCB-IR-theta=-1}) are given as: 
\begin{eqnarray}
\label{ci's-theta=-1-general}
& &  y_2=\frac{17}{210 t_0}, \ \ \ y_4=-\frac{597941}{120393000 t_0^3}.
\end{eqnarray}

\item $v_{2,4}$ appearing in (\ref{t[r]-soln-NCB-IR-theta=-1}) are given as: 
\begin{eqnarray}
\label{ci's-theta=-1}
& &  v_2=\frac{17}{210 t_0}, \ \ \ \  \ \ \ v_4=-\frac{289}{54600 t_0^3}.
\end{eqnarray}
\end{itemize}
In general, the coefficients in the series expansion
(\ref{ansatz-t[r]})\ (and similar other places in the paper), scale as
$c_n \sim \# \frac{1}{t_0^{n-1}}$\,, with ``$\#$'' is some numerical
coefficient that becomes an increasingly bigger (more unwieldy)
fraction at higher order $n$.

\section{EE, finite subregions ($A\neq 0$),
  $AdS_5$ Kasner}\label{FA-EE-AdS5K}
\setcounter{equation}{0}
\seceqcc

Here we give a brief description of the entangling RT/HRT surface for
finite subregions, \ie\ finite $A$, developing numerically the studies
in \cite{Manu:2020tty}.
The equation of motion for the entangling RT/HRT surface for finite $A$
in $AdS_{5}$ Kasner spacetime is given by (\ref{EOM-t[r]-EE}) with $d_i=3$.
The perturbative solution of this equation is the same as
(\ref{soln-t(r)-AdS5K-EE}) for $A=0$. For nonzero $A$, we solve
numerically for $t(r)$ up to the turning point $r_*$ determined by the
condition (see (\ref{EE-AdSK}), (\ref{l-scaling}))
\begin{eqnarray}\label{TP}
& & A=\phi_*=\frac{t(r_*)}{r_*^{d_i}}\,.
\end{eqnarray}
The perturbative solution (\ref{soln-t(r)-AdS5K-EE}) simplifies
(\ref{TP}) with $d_i=3$ to
\begin{equation}
6 A r_*^3 t_0-r_*^2-6 t_0^2=0\,.
\end{equation}
This can be solved for $r_*$ (with one real solution) but in
perturbation theory, it is consistent to take $r_*\sim At_0^{1/d_i}$
since $t(r_*)\sim t_0$, \ie\ the surface is approximately on the $t_0$
constant time slice (the surface bends very little, as we confirm below).
\begin{figure}[h]
\begin{subfigure}
  \centering
  \includegraphics[width=.5\linewidth]{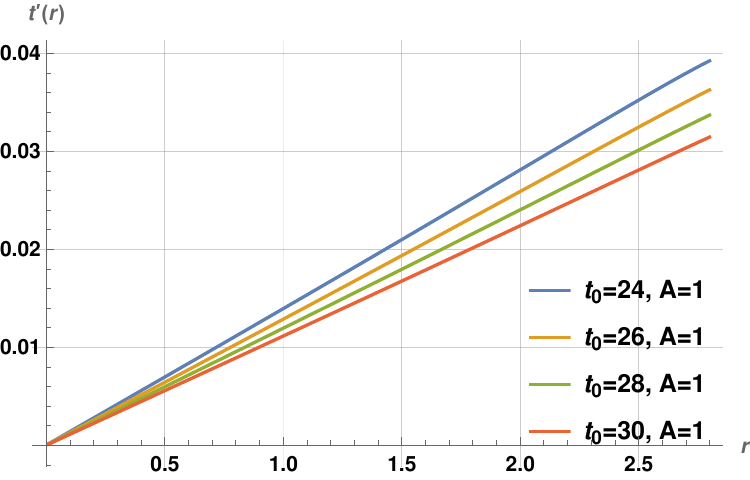}
\end{subfigure}
\begin{subfigure}
  \centering
  \includegraphics[width=.5\linewidth]{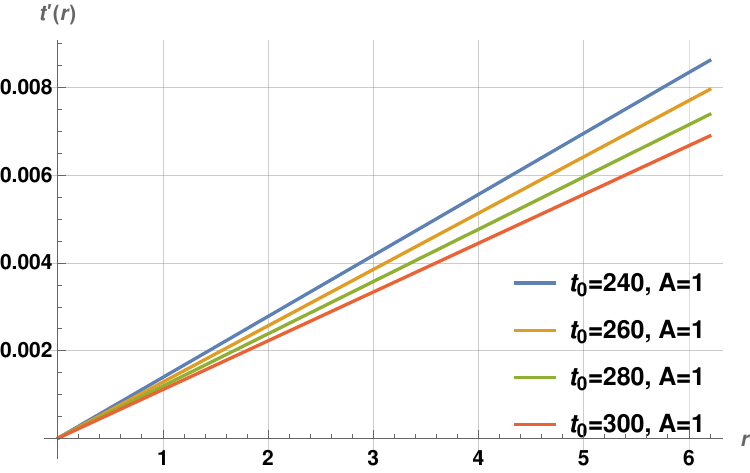}
\end{subfigure}
\caption{ Numerical plots of $t'(r)$ with $r$ in $AdS_5$ Kasner
   for different $t_0$ slices with $A=1$.}
\label{FA-AdS5K-EE}
\end{figure}
The $t(r)$-equation (\ref{EOM-t[r]-EE}) with $d_i=3$ is solved
numerically for the boundary conditions extracted from the perturbative
solution (\ref{soln-t(r)-AdS5K-EE}). The numerical solution for the
surface only makes sense upto the turning point $r_*$. We illustrate
this fixing $A=1$ so $r_*\sim t_0^{1/3}$ here, with the results plotted
in Fig.~\ref{FA-AdS5K-EE}.
It is clear that the bending is always small, \ie\ $t'(r)\ll 1$ over
the entire surface as expected: the $t'(r)$ values in
Fig.~\ref{FA-AdS5K-EE} are in approximate agreement with the
semiclassical $t'\sim {r\over 3t_0}$ in (\ref{soln-t(r)-AdS5K-EE}).
No lightlike limit arises here as expected (see Fig.~1 of
\cite{Goswami:2021ksw} for a qualitative picture of the surface).
This shows consistency of our techniques and analysis throughout
the paper where the numerics for complexity and entanglement for
large subregions with $A=0$ (Fig.~\ref{t[r]-plots-AdS5K-EE}) exhibits
clear lightlike limits.

\end{document}